\pdfoutput=1

\documentclass[superscriptaddress,pra,twocolumn,nofootinbib,notitlepage]{revtex4-1}

\usepackage{graphics,graphicx,amsthm,amsmath,amssymb}
\usepackage{color}
\usepackage{mathtools}
\usepackage{setspace}

\usepackage{float}

\usepackage[caption = false]{subfig}

\input{Qcircuit}

\def\({\left(}
\def\){\right)}

\def\bra#1{\mathinner{\langle{#1}|}}
\def\ket#1{\mathinner{|{#1}\rangle}}
\def\braket#1#2{\mathinner{\langle{#1}|#2 \rangle}}
\def\avg#1{\mathinner{\langle{#1} \rangle}}

\def\CZ{{\rm CZ}}

\def\T{{\rm T}}
\def\X{{\rm X}}
\def\Y{{\rm Y}}

\def\Z{{\rm Z}}
\def\H{{\rm H}}

\def\mE{{\mathbb E}}
\def\cK{{\mathcal K}}
\def\IPR{{\rm IPR}}
\def\poly{{\rm poly}}
\def\Pr{{\rm Pr}}
\def\exp{{\rm exp}}
\def\cl{{\rm pcl}}

\newcommand{\isep}{\mathrel{{.}\,{.}}\nobreak}

\begin{document}
\title{Characterizing Quantum Supremacy in Near-Term Devices}

\author{Sergio Boixo}
\affiliation{Google Inc., Venice, CA 90291, USA}
\author{Sergei V. Isakov}
\affiliation{Google Inc., 8002 Zurich, Switzerland}
\author{Vadim N. Smelyanskiy}
\affiliation{Google Inc., Venice, CA 90291, USA}
\author{Ryan Babbush}
\affiliation{Google Inc., Venice, CA 90291, USA}
\author{Nan Ding}
\affiliation{Google Inc., Venice, CA 90291, USA}
\author{Zhang Jiang}
\affiliation{QuAIL, NASA Ames Research Center, Moffett Field, CA 94035, USA}
\affiliation{SGT Inc., 7701 Greenbelt Rd., Suite 400, Greenbelt, MD 20770}
\author{Michael J. Bremner}
\affiliation{Centre for Quantum Computation and Communications Technology, Centre for Quantum Software and Information,  University of Technology Sydney, NSW 2007, Australia }
\author{John M. Martinis}
\affiliation{Google Inc., Santa Barbara, CA 93117, USA}
\affiliation{Department of Physics, University of California, Santa Barbara, CA 93106, USA}
\author{Hartmut Neven}
\affiliation{Google Inc., Venice, CA 90291, USA}

\date{\today}

\begin{abstract}
A critical question for the field of quantum computing in the near future is whether quantum devices without error correction can perform a well-defined computational task beyond the capabilities of state-of-the-art classical computers, achieving so-called quantum supremacy. We study the task of sampling from the output distributions of  (pseudo-)random quantum circuits, a natural task for benchmarking quantum computers. Crucially, sampling this distribution classically requires a direct numerical simulation of the circuit, with computational cost exponential in the number of qubits. This requirement is typical of chaotic systems. We extend previous results in computational complexity to argue more formally that this sampling task must take exponential time in a classical computer.
We study the convergence to the chaotic regime using extensive supercomputer simulations, modeling circuits with up to 42 qubits - the largest quantum circuits simulated to date for a computational task that approaches quantum supremacy.
We argue that while chaotic states are extremely sensitive to errors, quantum supremacy can be achieved in the near-term with approximately fifty superconducting qubits. 
We introduce cross entropy as a useful benchmark of quantum circuits which approximates the circuit fidelity. We show that the cross entropy can be efficiently measured when circuit simulations are available. Beyond the classically tractable regime, the cross entropy can be extrapolated and compared with theoretical estimates of circuit fidelity to define a  practical quantum supremacy test. 
\end{abstract}
\maketitle

\section{Introduction}\label{sec:intro}
Despite a century of research, there is no known method for efficiently simulating arbitrary quantum dynamics using classical computation. In practice, we are unable to directly simulate even modest depth quantum circuits acting on approximately fifty qubits. This strongly suggests that the controlled evolution of ideal quantum systems offers computational resources more powerful than classical
computers~\cite{feynman_simulating_1982,shor_algorithms_1994}. In this paper we build on existing results in quantum chaos~\cite{porter1956fluctuations,peres_stability_1984,schack_hypersensitivity_1993,beenakker1997random,emerson2003pseudo,scott_hypersensitivity_2006,gorin_dynamics_2006,dahlsten2007emergence,ambainis2007quantum,arnaud2008efficiency,trail_entanglement_2008,harrow2009random,weinstein_parameters_2008,brown2010convergence,brown_scrambling_2012,PhysRevLett.111.127205,hosur2015chaos} and computational complexity theory~\cite{aaronson2003quantum,terhal2004adaptive,aaronson2005quantum,bremner_classical_2011,aaronson2011computational,fujii2013quantum,aaronson2014equivalence,fujii2014impossibility,jozsa_classical_2014,bremner2015average,farhi_quantum_2016} to propose an experiment for characterizing ``quantum supremacy''~\cite{preskill_2012} in the presence of errors. We study the computational task of sampling from the output distribution of random quantum circuits composed from a universal gate set, a natural task for benchmarking quantum computers.  We propose the cross entropy difference as a measure of correspondence between experimentally obtained samples and the output distribution of the ideal circuit. Finally, we discuss a robust set of conditions which should be met in order to be sufficiently confident that an experimental demonstration has actually achieved quantum supremacy. Quantum supremacy is achieved when a formal computational task is performed with an existing quantum device which cannot be performed using any known algorithm running on an existing classical supercomputer in a reasonable amount of time.

In this paper we show how to estimate the cross entropy between an experimental implementation of a random quantum circuit and the ideal output distribution simulated by a supercomputer. We study numerically the convergence of the output distribution to the Porter-Thomas distribution, characteristic of quantum chaos. We find a good convergence for the first ten moments and the entropy at depth 25 with circuits of up to $7 \times 6$ qubits in a 2D lattice. Using chaos theory, the properties of the Porter-Thomas distribution, and numerical simulations, we argue that the cross entropy is closely related to the circuit fidelity. State-of-the-art supercomputers cannot simulate universal random circuits of sufficient depth in a 2D lattice of approximately $7 \times 7$ qubits with any known algorithm and significant fidelity.

Time accurate simulations of classical dynamical systems with chaotic behavior are among the hardest numerical tasks. Examples include turbulence and population dynamics, essential for the study of meteorology, biology, finance, etc. 
In all these cases, a direct numerical simulation is required in order to get an accurate description of the system state after a finite time. A signature of chaotic systems is that small changes in the model specification lead to large divergences in system trajectories. This phenomenon is described by Lyapunov exponents and generally requires computational resources that grow exponentially in time.  

While we do not provide a formal definition of quantum chaos here, we review several well known characteristics of quantum chaos to argue that sampling the output distribution of a random quantum circuit is a hard computational task. In analogy with classical Lyapunov exponents, a signature of quantum chaos is the  decrease of the overlap $|\braket{\psi_t}{\psi_t^\epsilon}|^ 2$ of the quantum state $\ket {\psi_t}$ with the state $\ket {\psi_t^\epsilon}$ resulting from a small  perturbation  $\epsilon$  to the Hamiltonian that evolves $\ket{\psi_t}$~\cite{peres_stability_1984,schack_hypersensitivity_1993,scott_hypersensitivity_2006,gorin_dynamics_2006}.  The overlap decreases exponentially in the evolution time $t$ and $\epsilon$ because chaotic evolutions give rise to delocalization of quantum states~\cite{beenakker1997random,emerson2003pseudo}. Such states are closely related to ensembles of random unitary matrices studied in random matrix theory~\cite{beenakker1997random,mehta2004random}, they possess no symmetries, and are spread over Hilbert space.
Therefore, as in the case of classical chaos, obtaining a description of $\ket{\psi_t}$ requires a high fidelity classical simulation. This challenge is compounded by the exponential growth of Hilbert space $N = 2^n$ with the qubit dimension $n$.

It follows that unless a classical algorithm uses resources that grow exponentially in $n$, its output would be almost statistically uncorrelated with the output distribution corresponding to general global measurements of the chaotic quantum state.\footnote{A classical algorithm that uses time and space resources that grow exponentially in $n$ can reconstruct all measurements of the chaotic quantum state exactly.}
Indeed, it has been argued that
classically solving related sampling problems requires computational resources with asymptotic exponential scaling~\cite{aaronson2003quantum,terhal2004adaptive,aaronson2005quantum,bremner_classical_2011,aaronson2011computational,fujii2013quantum,aaronson2014equivalence,fujii2014impossibility,jozsa_classical_2014,bremner2015average,farhi_quantum_2016}. Examples include BosonSampling~\cite{aaronson2011computational} and  approximate simulation of commuting quantum computations~\cite{bremner_classical_2011,bremner2015average}.

\begin{figure}
  \centering
  
\begin{align*}
  \Qcircuit @C=.5em @R=.5em { 
\lstick{\ket{0}}  &\gate{\H} 
  & \ctrl{1} & \gate{\T} & \qw       & \ctrl{1}  & \gate{\X^{1/2}} & \qw & \qw  {/} & {/} & \qw & \qw & \meter \\ 
 \lstick{\ket{0}} &\gate{\H} & \ctrl{0} & \ctrl{1}  & \gate{\T} & \ctrl{0}  & \ctrl{1}       & \qw & \qw  {/} & {/} & \qw & \gate{\Y^{1/2}} &\meter \\ 
 \lstick{\ket{0}} &\gate{\H} & \qw      & \ctrl{0}  & \ctrl{1}  & \gate{\T} & \ctrl{0}       & \qw & \qw  {/} & {/} & \qw & \ctrl{1} &\meter \\ 
 \lstick{\ket{0}} &\gate{\H} & \ctrl{1} & \gate{\T} & \ctrl{0}  & \ctrl{1}  & \gate{\Y^{1/2}} & \qw & \qw  {/} & {/} & \qw & \ctrl{0} &\meter \\ 
 \lstick{\ket{0}} &\gate{\H} & \ctrl{0} & \gate{\T} & \qw       & \ctrl{0}  & \qw            & \qw & \qw  {/} & {/ }& \qw & \qw &\meter
} 
\end{align*}
  \caption{Example of a random quantum circuit in a 1D array of qubits. Vertical lines correspond to controlled-phase ($\CZ$) gates (see Sec.~\ref{sec:pt}). }\label{fig:1dc}
\end{figure}
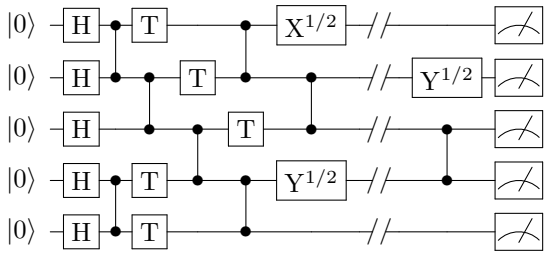

Random quantum circuits with gates sampled from a universal gate set are examples of quantum chaotic evolutions that naturally lend themselves to the quantum computational framework~\cite{emerson2003pseudo,dahlsten2007emergence,ambainis2007quantum,arnaud2008efficiency,harrow2009random,brown2010convergence}. A circuit, corresponding to a unitary transformation $U$, is a sequence of $d$ clock cycles of one- and two-qubit gates, with gates applied to different qubits in the same cycle, see Fig.~\ref{fig:1dc}. With realistic superconducting hardware constraints~\cite{barends_superconducting_2014,kelly_state_2015}, gates act in parallel on distinct sets of qubits restricted to a 1D or 2D lattice. 

In this paper we study the computational task of sampling bit-strings from the distribution defined by the output state $\ket \psi$ of a (pseudo-)random quantum circuit $U$ of size \emph{polynomial} in $n$. We will compare the sampling output of $U$ to a generic classical sampling algorithm that takes a specification of  $U$ as input and samples a bit-string with computational time cost also \emph{polynomial} in $n$. We will show that a bit-string sampled from $U$ is typically $e$ times more likely than a bit-string sampled by the classical algorithm. 
A quantum sample $S$ of $m$ measurement outcomes $x \in \{0,1\}^n$ in a local qubit basis has probability 
$\Pi_{x \in S} |\braket {x} {\psi}|^2 $. Denote by $S_\cl$ a sample of $m$ bit-strings from the polynomial classical algorithm. We argued above standard assumptions in chaos theory that in this case $S_\cl$ is expected to be almost uncorrelated with the distribution defined by $\ket \psi$. We will substantiate this numerically and theoretically in later sections. The sample $S_\cl$ is assigned a probability $\Pi_{x \in S_\cl} |\braket {x} {\psi}|^2$ by the distribution defined by $\ket \psi$. As we show in this paper, the ratio of these probabilities for a sufficiently large circuit in the typical case is, within logarithmic equivalence,  $\Pi_{x \in S} |\braket {x} {\psi}|^2  / \Pi_{x \in S_\cl} |\braket {x} {\psi}|^2 \sim e^{m}$ (see Eq.~\eqref{eq:lr}). 
We will also show that for a typical sample $S_\exp$ produced by an experimental implementation of $U$ this ratio is, within logarithmic equivalence,
\begin{align}
  \label{eq:11}
  \frac {\Pi_{x \in S_\exp} |\braket {x} {\psi}|^2}{ \Pi_{x \in S_\cl} |\braket {x} {\psi}|^2} \sim e^{m \,e^{\!-r g}} \gg 1\;,
\end{align}
where the parameter $r$ provides an estimate of the effective per-gate error rate, and $g \propto n d$ is the total number of gates (see Eqs.~\eqref{eq:ae} and \eqref{eq:aap}).
Note the double exponential structure in Eq.~\eqref{eq:11} with two large parameters $m,g \gg 1$. Therefore, the ratio of probabilities in Eq.~\eqref{eq:11}, an experimentally observable quantity, is enormously sensitive to the effective per-gate error rate $r$. The parameter $r$ can serve as an extremely accurate characterization of the degree of correlation of $S_\exp$ with the distribution defined by $U$, and provides a novel tool for benchmarking complex multiqubit quantum circuits.
We will argue that $r$ can be estimated theoretically and compared with experiments to define a quantum supremacy test.

We now give the main outline of the paper. In Sec.~\ref{sec:cqs} we obtain Eq.~\eqref{eq:11} from the cross entropy between the two distributions and we explain how it can be measured in an experiment. In Sec.~\ref{sec:fid} we explain theoretically and numerically why the cross entropy is closely related to the overall circuit fidelity. We also introduce an effective error model for the overall circuit, and compare it with numerical simulations of the circuit with digital errors. In Sec.~\ref{sec:pt} we study numerically the convergence of the circuit output to the Porter-Thomas distribution, characteristic of quantum chaos. In Sec.~\ref{sec:ch} we use complexity theory to argue that this sampling problem is computational hard.

\section{Characterizing quantum supremacy}\label{sec:cqs}

\subsection{Ideal circuit vs. polynomial classical algorithm}

\begin{figure}
  \centering
  \includegraphics[width=\columnwidth]{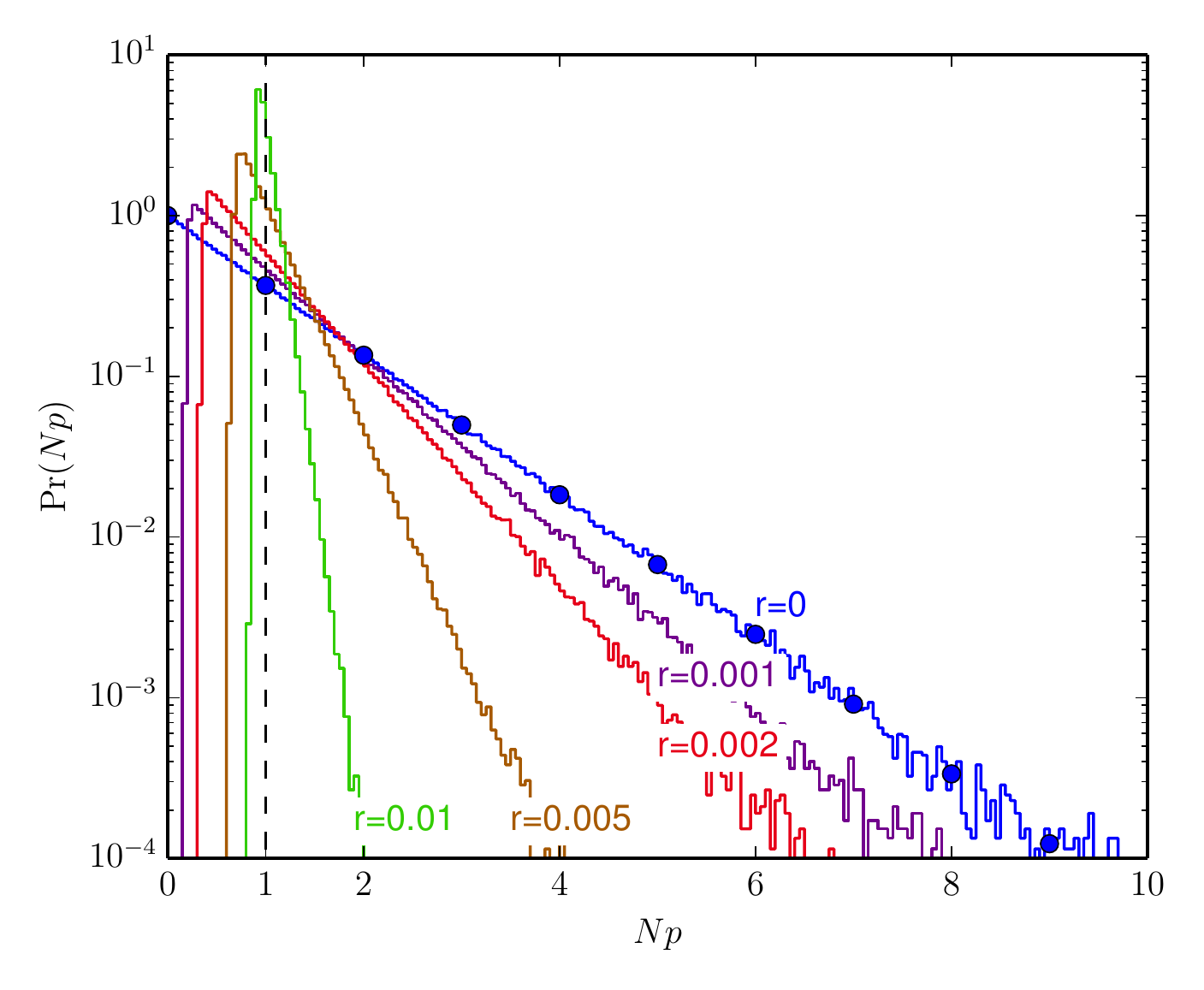}
  \caption{Distribution function of rescaled probabilities $Np$ to observe individual bit-strings as an output of a typical random circuit. Blue curve ($r=0$) shows the distribution of  $\{N p_U(x_j)\}$ obtained from numerical simulations of the ideal random circuit  (see Sec.~\ref{sec:pt}) . This distribution is very close to the Porter-Thomas form $\Pr(Np) = e^{-N p}$ shown with blue dots. Curves with different colors show the distributions of probabilities obtained for different Pauli error rates $r$. The dashed line at $Np = 1$ corresponds to the uniform distribution $\delta(p-1/N)$. These numerics are obtained from simulations of a planar circuit with $5 \times 4$ qubits and gate depth of 40 ($n=20$ and $N = 2^{20}$). }
  \label{fig:pnp}
\end{figure}

Consider a state $\ket{\psi_d}$ produced by a random quantum circuit. Due to delocalization, the real and imaginary parts of the amplitudes $\braket {x_j} {\psi_d}$ in any local qubit basis $\{x_j\}_{j=1}^N$, $x_j \in \{0,1\}^n$ are approximately uniformly distributed in a $2N=2^{n+1}$ dimensional sphere (Hilbert space) subject to the normalization constraint. This implies that their distribution is an unbiased Gaussian with variance $\propto 1/N$, up to finite moments~\cite{renes_symmetric_2004}. This distribution is a signature of delocalization due to quantum correlations manifested as level repulsion in systems with stationary Hamiltonians. The distribution of measurement probabilities $p(x_j) = |\braket {x_j} {\psi_d}|^2$  approaches the exponential form $N e^{-N p}$, known as Porter-Thomas~\cite{porter1956fluctuations}, see Fig.~\ref{fig:pnp}. The probability vectors thus obtained are uniformly distributed  over the probability simplex (i.e., according to the symmetric Dirichlet distribution).

The circuit depth or time to approach the Porter-Thomas regime is expected to correspond to the ballistic spread of entanglement across Hilbert space in chaotic systems~\cite{PhysRevLett.111.127205,hosur2015chaos}. This timescale grows as $n^{1/D}$ where $D$ is the dimension of the qubit lattice. In particular, $D=1$ for a linear array~\cite{brandao_local_2012,nakata2016efficient}, $D=2$ for a square lattice~\cite{brown_scrambling_2012}, and $D$ goes to infinity for a fully connected architecture~\cite{harrow2009random,weinstein_parameters_2008,brown_scrambling_2012} (see Sec.~\ref{sec:pt}). 

The output probability $p(x_j)$ of each bit-string from a random quantum circuit is of order $1/N = 2^{-n}$, see Fig.~\ref{fig:pnp}. Therefore,  each bit-string in a sample of size polynomial in $n$ will be unique. In other words, the output of a random quantum circuit can not be distinguished from a uniform sampler over  $\{x_j\}$ unless we pre-compute the specific output probabilities $p(x_j)$~\cite{lloyd2013pure,popescu2006entanglement,gogolin_boson-sampling_2013,ududec_information-theoretic_2013}.\footnote{In the case of BosonSampling, generic observables sensitive to Boson statistics can be used to distinguish the output distribution from uniform~\cite{aaronson2014bosonsampling,walschaers2016statistical}. Nevertheless, it is also unlikely that a Bosonsampler can be distinguished from classically efficient simulations unless we use exponential resources~\cite{aaronson2011computational,aaronson2014bosonsampling}.}

Nevertheless, the Porter-Thomas distribution  $N e^{-Np}$ has substantial support on values $Np < 1$, see Fig.~\ref{fig:pnp}. This will allow us to clearly distinguish  it from the uniform distribution over $\{x_j\}$, which has a form given by a delta function $\delta(p-1/N)$, after computing $p(x_j)$ with a powerful enough classical computer.  
 Circuit specific global measurements can be sensitive to time-accurate simulations of chaotic quantum state evolutions.\footnote{Specifically, the $\ell_1$ norm distance between the Porter-Thomas distribution and the uniform distribution over $\{x_j\}$ is $2/e$, independent of $n$. Therefore, information theoretically, a constant small number of measurements are sufficient to distinguish these distributions.} Therefore, such observables will be extremely hard to simulate classically. 

Let $\ket{\psi}=U\ket{\psi_0}$ be the output of a given random circuit $U$. Consider a sample $S = \{x_{1},\ldots, x_{m}\}$ of bit-strings $x_j$ obtained from $m$ global measurements of every qubit in the computational basis $\{\ket{x_j}\}$ (or any other basis obtained from local operations). The joint probability of the set of outcomes   $S$ is $\Pr_U(S)=\prod_{x_j \in S}p_U(x_j)$ where $p_U(x)$ $\equiv$ $|\braket {x} {\psi}|^2$. For a typical sample $S$, the central limit theorem implies that
\begin{align}
  \log \Pr_U(S) & = \sum_{x_j \in S} \log p_U(x_j) \nonumber \\ &=  - m\,\H(p_U) + O(m^{1/2}) \;,\label{eq:lps}
\end{align}
where $\H(p_U)$ $\equiv$ $- \sum_{j=1}^N p_U(x_j) \log p_U(x_j) $ is the entropy of the output of $U$. Because $p_U(x)$ are approximately {\it i.i.d.} distributed according to the Porter-Thomas distribution, if follows that 
\begin{align}
\H(p_U)&= - \int_0^\infty p N^2 e^{-Np} \log p \,dp \nonumber \\ &= \log N -1 +\gamma\;,\label{eq:eu}
\end{align}
where $\gamma \approx 0.577$ is the Euler constant.

 Let $A_\cl(U)$ be a classical algorithm with computational time cost \emph{polynomial} in $n$ that takes a specification of the random circuit $U$ as input and outputs a bit-string $x$ with probability distribution $p_\cl(x|U)$. Consider a typical sample $S_\cl = \{x_{1}^\cl,\ldots,x_{m}^\cl\}$ obtained from $A_\cl(U)$.  We now focus on the probability $\Pr_U(S_\cl)=\prod_{x_j^\cl \in S_\cl} p_U(x_j^\cl)$  that this sample  $S_\cl$ is observed from the output $\ket{\psi}$ of the circuit $U$. The central limit theorem implies that
\begin{align}
  \log \Pr_U(S_\cl) =  - m\,\H(p_\cl,p_U) + O(m^{1/2}) \;,\label{eq:lpsu}
\end{align}
where 
\begin{align}
  \label{eq:15}
  \H(p_\cl,p_U) \equiv - \sum_{j=1}^N p_\cl(x_j|U) \log p_U(x_j)
\end{align}
is the cross entropy between $p_\cl(x|U)$ and $p_U(x)$. Note that if the cross entropy $\H(p_\cl,p_U)$ is larger than the entropy $\H(p_U)$, this implies that $p_\cl(x|U)$ is sampling bit-strings that have lower probability of being observed by the circuit $U$. 

We are interested in the average quality of the classical algorithm. Therefore, we average the cross entropy over an ensemble $\{U\}$ of random circuits 
\begin{align}
  \label{eq:Uens}
  \mE_U\left[\H(p_\cl,p_U)\right] &= \mE_U\left[\sum_{j=1}^N p_\cl(x_j |U)\log {1 \over  p_U(x_j)} \right].
\end{align}
We will give numerical evidence in Secs.~\ref{sec:fid} (see also Apps.~\ref{app:correlations} and ~\ref{app:feynman}), and computational complexity theory arguments in Sec.~\ref{sec:ch}, that a direct numerical simulation of the evolution is required in order to get an accurate description of the system state after a finite time. Therefore, consistent with aforementioned insights from quantum chaos, we assume that the output of a classical algorithm with polynomial cost  is almost statistically uncorrelated with $p_U(x)$. In particular, as we will show numerically in Secs.~\ref{sec:fid} and~\ref{sec:pt}, and in App.~\ref{app:feynman}, a direct numerical simulation of the evolution is required in order to get an accurate description of the system state after a finite time. 

Thus, averaging over the ensemble $\{U\}$ can be done independently for the output of the polynomial classical algorithm $p_\cl(x|U)$ and $\log p_U(x)$. The distribution of universal random quantum circuits converges to the uniform (Haar) measure with increasing depth~\cite{emerson2003pseudo,emerson_convergence_2005,harrow2009random}. For fixed $x_j$, the distribution of values $\{p_U(x_j)\}$ when unitaries are sampled from the Haar measure also has the Porter-Thomas form. Therefore, we assume that we use random circuits of sufficient depth such that
\begin{align}
  - \mE_U\left[\log p_U(x_j) \right]&\approx - \int_0^\infty  N e^{-Np} \log p \,dp \nonumber \\&= \log N +\gamma\;.\label{eq:eue}
\end{align}
Note that this equation is similar to Eq.~\eqref{eq:eu}, except that the integrand here is missing a factor of $Np$. Then using $\sum_{j=1}^N p_\cl (x_j|U) =1$ we get
\begin{align}
  \label{eq:au}
  \mE_U\left[\H(p_\cl,p_U)\right] &= \log N +\gamma  \;.
\end{align}
From Eqs.~(\ref{eq:lps}-\ref{eq:eu}) and (\ref{eq:lpsu}-\ref{eq:au}) we obtain
\begin{align}
  \mE_U\left[{\log \Pr_U(S) - \log  \Pr_U(S_\cl)} \right] &\simeq m\;.\label{eq:lr}
\end{align}
Equation~\eqref{eq:lr} reveals the remarkable property that a typical sample $S$ from a random circuit $U$ represents a signature of that circuit.  Note that the l.h.s. is the expectation value of the log of $\Pi_{x \in S} |\braket {x} {\psi}|^2  / \Pi_{x^\cl \in S_\cl} |\braket {x^\cl} {\psi}|^2$. The  numerator is dominated by measurement outcomes $x$ that have  high measurement probabilities $|\braket {x} {\psi}|^2 > 1/N$. Conversely, the values of $x^\cl$ in the denominator are essentially uncorrelated with the output distribution of $U$. Therefore, they are dominated by the support of the Porter-Thomas distribution with $p < 1/N$.

\subsection{Cross entropy difference}

We note that the result in Eq.~\eqref{eq:au} also corresponds to the cross entropy $\H_0 = \log N + \gamma$ of an algorithm which picks bit-strings uniformly at random, $p_0(x) = 1/N$.  This leads to a proposal for a test of quantum supremacy.
We will measure the quality of an algorithm $A$ for a given number of qubits $n$ as the difference between its cross entropy and the cross entropy of a uniform classical sampler. The algorithm $A$ can be an experimental quantum implementation, or a classical algorithm implementation with \emph{polynomial} or \emph{exponential} cost as long as it is actually executed on an existing classical computer. We call this quantity the cross entropy difference:
\begin{align}
  \label{eq:5}
  \Delta\H(p_A) &\equiv \H_0 - \H(p_A,p_U ) \nonumber \\ &=\sum_j \(\frac 1 N - p_A (x_j|U)  \){\log \frac 1 {p_U(x_j)}}\;.
\end{align}
The cross entropy difference measures how well algorithm $A(U)$ can predict the output of a (typical) quantum random circuit $U$. This quantity is unity for the ideal random circuit if the entropy of the output distribution is equal to the entropy of the Porter-Thomas distribution, and zero for the uniform distribution, see Eqs.~\eqref{eq:eu} and~\eqref{eq:au}. 

In an experimental setting we describe the evolution of the density matrix
\begin{align}
\rho_\cK= \cK_U(\ket{\psi_0}\!\bra{\psi_0})\label{eq:12}
\end{align}
 with a superoperator $\cK_U$ which corresponds to the circuit $U$ and takes into account initialization, measurement and gate errors. We refer to the experimental implementation as $A_\exp(U)$ and associate with it the probability distribution $p_\exp(x_j|U) = \bra{x_j} \rho_\cK \ket{x_j}$ and sample $S_\exp$. Consistent with Eq.~\eqref{eq:11}, the experimental cross entropy difference is
 \begin{align*}
   \alpha \equiv \mE_U[\Delta \H(p_\exp)]\;.
 \end{align*}
Quantum supremacy is achieved, in practice, when
\begin{align}
  \label{eq:1ac}
  1 \ge \alpha > C\;,
\end{align}
where a lower bound for $C$ (see also discussion below) is given by the performance of the best classical algorithm $A^*$ known executed on an existing classical computer, 
\begin{align}\label{eq:c}
  C = \mE_U[\Delta \H(p^*)  ]\;.
\end{align}
Here $p^*$ is the output distribution of $A^*$.

The space and time complexity of simulating a random circuit by using tensor contractions is exponential in the treewidth of the quantum circuit, which is proportional to $\min(d,n)$ in a 1D lattice, and $\min(d\sqrt n, n)$ in a 2D lattice~\cite{markov_simulating_2008,aaronson2016complexity}. For large depth $d$, algorithms are limited by the memory required to store the wavefunction in random-access memory, which in single precision is $2^n \times 2 \times  4$ bytes. For $n=48$ qubits this requires at least 2.252 Petabytes, which is approximately the limit of what can be done on today’s large-scale supercomputers.\footnote{ Trinity, the sixth fastest supercomputer in TOP500~\cite{top500}, has $\sim 2$ Petabytes of main memory –- one of the largest among existing supercomputers today.} For circuits of small depth or less than approximately 48 qubits, direct simulation is viable so $C=1$ and quantum supremacy is impossible. Beyond this regime we are limited to an estimation of the Feynman path integral corresponding to the unitary transformation $U$. In this regime, the lower bound for $C$ decreases exponentially with the number of gates $g \gg n$, see App.~\ref{app:feynman}.

We now address the question of how the cross entropy difference $\alpha$ can be estimated from an experimental sample of bit-strings $S_\exp$ obtained by measuring the output of $A_\exp(U)$ after $m$ realizations of the circuit. For a typical sample $S_\exp$, the central limit theorem applied to Eq.~\eqref{eq:5} implies that
\begin{align}
  \label{eq:ae}
  \alpha \simeq \H_0  - \frac 1 m \sum_{j =1}^{m} {\log \frac 1 {p_U(x_j^\exp )}}\;,
\end{align}
where $\H_0$ is defined after Eq.~\eqref{eq:au}. The statistical error in this equation, from the central limit theorem, goes like $ \kappa / \sqrt m$, with $\kappa \simeq 1$. The experimental estimation would proceed as follows:
\begin{enumerate}
\item Select a random circuit $U$ by sampling from an available universal set of one and two-qubit gates, subject to experimental layout constraints. 
\item Take a sufficiently large sample $S_\exp=\{x_{1}^\exp,\ldots,x_{m}^\exp\}$ of bit-strings $x$ in the computational basis ($m \sim 10^3 - 10^6$).
\item Compute the quantities $\log 1/ p_U(x_j^\exp)$  with the aid of a sufficiently powerful classical computer.
\item Estimate $\alpha$ using Eq.~\eqref{eq:ae}. 
\end{enumerate}

For large enough circuits, the quantity $p_U(x_j^\exp)$ can no longer be obtained numerically. 
At this point, $C \simeq 0$, and supremacy can be achieved. Unfortunately, this also implies that $\alpha$ can no longer be measured directly. We argue that the observation of a close correspondence between experiment, numerics and theory would provide a reliable foundation from which to extrapolate $\alpha$.  The value of $\alpha$ can be extrapolated from circuits that can be simulated because they have either less qubits (direct simulation), mostly Clifford gates (stabilizer simulations)~\cite{bravyi_improved_2016} or smaller depth (tensor contraction simulations)~\cite{markov_simulating_2008,aaronson2016complexity}.

In practice, the necessary value of $\alpha$ in Eq.~\eqref{eq:1ac} to claim quantum supremacy will be limited not only by the lower bound on $C$ in Eq.~\eqref{eq:c}, but also by the number of measurements necessary to estimate $\alpha$ with high precision in Eq.~\eqref{eq:ae}, possible experimental biases among the different circuit types used to extrapolate $\alpha$, and the precision in the agreement between theory and experiment.
Next, we present a theoretical error model for $\cK_U$ (see Eq.~\eqref{eq:12}) and the corresponding estimate of $\alpha$ that can be compared with experiments.

\section{Fidelity analysis}\label{sec:fid}

\begin{figure}
  \centering
  \includegraphics[width=\columnwidth]{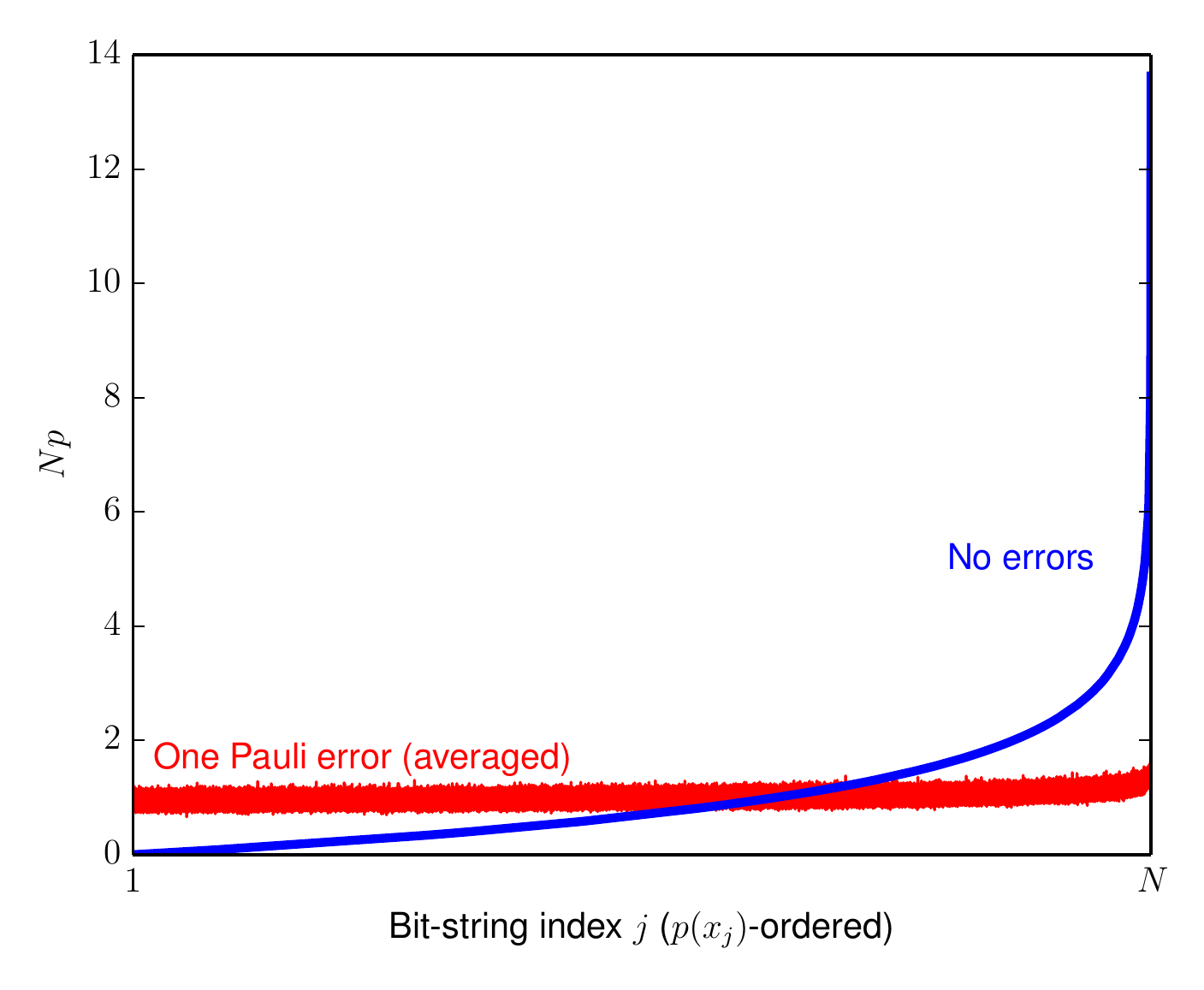}
  \caption{The blue line shows the probabilities $p_U(x_j)$ of bit-strings $x_j$ sorted in ascending order. The red line shows the corresponding probabilities after adding a Pauli error ($X$ or $Z$) in a single location in the circuit, using the same ordering. The circuit used has $5 \times 4$ qubits and depth 40 (see Sec.~\ref{sec:pt}). We average over all possible error locations. The average over errors gives almost the uniform distribution. The small residual correlation (slight upper curvature seen in the red line) is analyzed numerically in App.~\ref{app:correlations}.}
  \label{fig:1error}
\end{figure}

The output $\rho_\cK$ of the experimental realization $\cK_U$ of a random circuit $U$ is
\begin{align}
  \label{eq:1}
  \rho_\cK = \tilde \alpha_\cK U \ket {\psi_0} \!\bra {\psi_0} U^\dagger + (1-\tilde\alpha_\cK) \sigma_\cK\;,
\end{align}
where $\bra  {\psi_0}U^\dagger\sigma_\cK U\ket {\psi_0}=0$, $\tilde \alpha_\cK$ is the circuit fidelity, and we assume incoherent errors. The density matrix $\sigma_\cK$ represents the effect of errors. The corresponding average cross entropy difference is
\begin{align}
  \alpha &= \mE_U[\H_0 + \sum_j \bra{x_j} \rho_\cK \ket{x_j} \log p_U(x_j)] \\
  &= \tilde \alpha +(1-\tilde \alpha) \H_0 \\ \nonumber&\quad\quad+ \mE_U\left[(1-\tilde \alpha_\cK)  \sum_j\bra{x_j}  \sigma_\cK \ket{x_j}  \log p_U(x_j)\right] \;,
\end{align}
where $\tilde \alpha = \mE_U[\tilde \alpha_\cK]$ is the average fidelity over random circuits and we used Eq.~\eqref{eq:eu}.

Because $U$ is a random circuit implementing a chaotic evolution, we see in numerical simulations (see Fig.~\ref{fig:1error} and App.~\ref{app:correlations}) that the probabilities $p_U(x)$ and $\bra x \sigma_\cK \ket x$ are almost uncorrelated.  Under this ansatz, by the same arguments leading to Eq.~\eqref{eq:au}, we obtain that the circuit fidelity $\tilde \alpha_\cK$ is approximately equal to the average cross entropy difference $\alpha$
\begin{align}
  \alpha =   \mE_U[\Delta \H(p_\exp)] \approx \tilde \alpha \;.\label{eq:aap}
\end{align}
Estimating the circuit fidelity by directly measuring the cross entropy (see Eq.~\eqref{eq:ae}) is a fundamentally new way to characterize complex quantum circuits. A similar result is obtained with coherent errors, although they will result in larger fluctuations around the mean. 

 The standard approach to studying circuit fidelity is the digital error model where each quantum gate is followed by an error channel~\cite{barends_digital_2015,knill2008randomized}. Within this model, the circuit fidelity can be estimated as~\cite{carlo2004simulating,barends_digital_2015}
\begin{align}
  \label{eq:6}
  \alpha \approx \exp(-r_1 g_1-r_2g_2-r_{\rm init}n-r_{\rm mes}n)\;,
\end{align}
where $r_1, r_2 \ll 1$ are the Pauli error rates  for one and two-qubit gates, $r_{\rm init}, r_{\rm mes}\ll 1$ are the initialization and measurement error rates, and $g_1, g_2 \gg 1$ are the numbers of one and two-qubit gates respectively.

\begin{figure}
  \centering
  \includegraphics[width=\columnwidth]{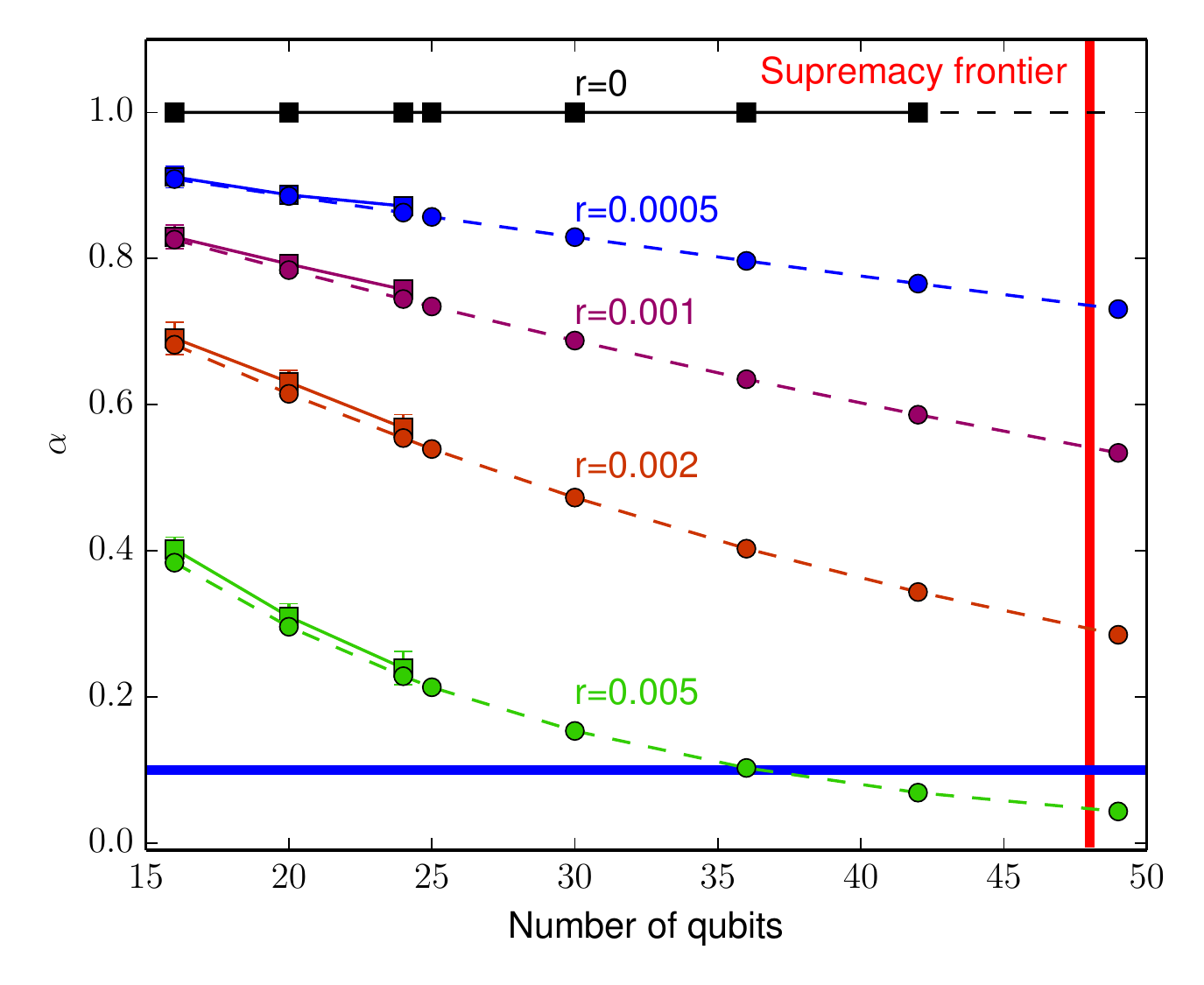}
  \caption{The circuit fidelity $\alpha$ as a function of the number of qubits. Different colors correspond to different Pauli error rates $r_2  = r_{\rm init} = r_{\rm mes} = r$ and $r_1 = r/10$. Circular markers correspond to the numerically simulated fidelities, Eq.~\eqref{eq:6}. Square markers correspond to the average cross entropy difference among 10 instances, Eq.~\eqref{eq:5}.  The circuit depth in these simulations is 40 (see Sec.~\ref{sec:pt}). The red line, at 48 qubits, is a reasonable estimate of the largest size that can be simulated with state-of-the-art classical supercomputers in practice. Using state-of-the-art superconducting circuits we expect $\alpha \gtrsim 0.1$ (blue line) for a $7 \times 7$  circuit. Error bars correspond to the standard deviation among instances. }
  \label{fig:sva}
\end{figure}

We have performed numerical simulations of random circuits in the presence of errors by introducing a depolarizing channel after each gate~\cite{barends_superconducting_2014,barends_digital_2015,kelly_state_2015,barends_digitized_2015,emerson_scalable_2005,knill2008randomized,magesan_robust_2011,magesan_characterizing_2012} (see Sec.~\ref{sec:pt} for details about the circuits design). Errors in the depolarizing channel after each two-qubit gate are emulated by applying one of the 15 possible combinations of products of two Pauli operators (excluding the identity) with an equal probability of $r_2/15$. Similarly, we apply a randomly selected single Pauli matrix after each one-qubit gate with an equal probability of $r_1/3$. Initialization and measurement errors are simulated by applying a bit-flip with probability $r_{\rm init}$ and  $r_{\rm mes}$ respectively. Figure~\ref{fig:sva} shows the cross entropy difference, Eq.~\eqref{eq:5}, obtained from these simulations, and the estimated fidelity, Eq.~\eqref{eq:6}. We observe a good agreement between these two quantities. The small difference between the cross entropy difference and the estimated fidelity is due to residual correlations analyzed numerically in App.~\ref{app:correlations}.

Note that the cross entropy difference of the ideal circuit ($r=0$ in the figure) is almost exactly one, indicating that at this depth all sizes studied are in the Porter-Thomas regime. Details of the optimizations employed for the simulation of the larger circuits, of up to $42$ qubits, are given in App.~\ref{app:qsd}. These are the largest quantum circuits simulated to-date for a computational task that approaches quantum supremacy. 

\begin{figure}
  \centering
  \includegraphics[width=\columnwidth]{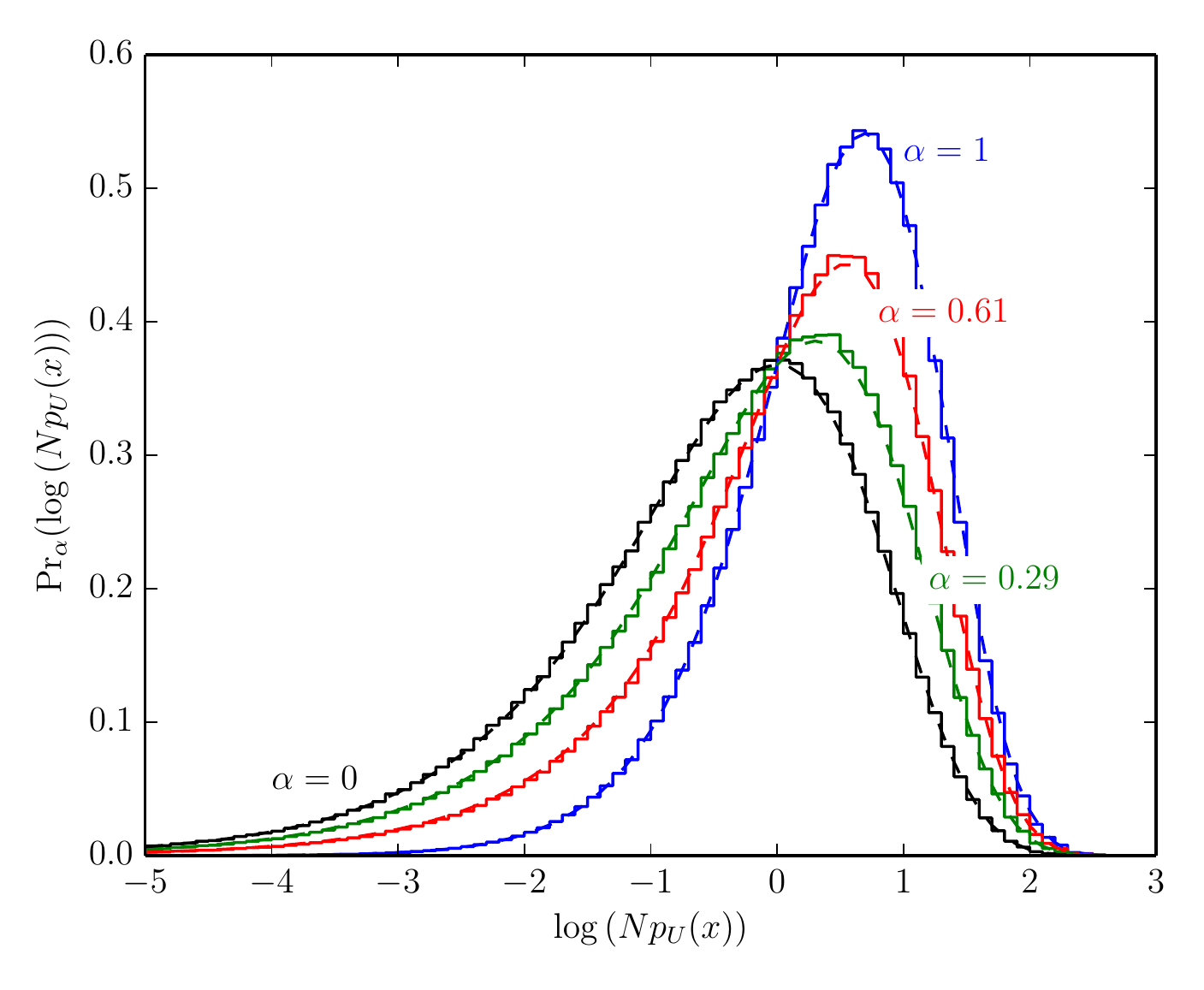}
  \caption{Probability distribution of $\log(Np_U(x))$ where bit-strings $x$ are sampled from a circuit of fidelity $\alpha$. The continuous step histograms are obtained from numerical simulations with different  Pauli error rates $r_2  = r_{\rm init} = r_{\rm mes} = r$ and $r_1 = r/10$. The values of $r$ are $r=0$ for $\alpha=1$ (blue), $r=0.005$ for $\alpha=0.43$ (red), $r=0.01$ for $\alpha=0.18$ (green) and uniform sampling of bit-strings for $\alpha=0$. The value of $\alpha$ is estimated using Eq.~\eqref{eq:6}. The superimposed dashed lines correspond to the theoretical distribution of Eq.~\eqref{eq:pal}. We chose a circuit of $5 \times 4$ qubits and depth 40 (see Sec.~\ref{sec:pt}).}
  \label{fig:pal}
\end{figure}

Because chaotic states are maximally entangled~\cite{trail_entanglement_2008,PhysRevLett.111.127205,hosur2015chaos,nahum_quantum_2016}, even one Pauli error completely destroys the state~\cite{boixo_operational_2008}, as seen in numerical data in  Fig.~\ref{fig:1error}. More formally, consider a sequence of arbitrary quantum channels interleaved with unitaries randomly chosen from a group that is also a 2-design. This is equivalent to a sequence of channels with the same average fidelity in which all the channels (except the last one) are transformed into depolarizing channels~\cite{emerson_scalable_2005,magesan_characterizing_2012}. 
Although individual two-qubit gates are not a 2-design for $n$ qubits, a large part of the evolution of a typical random circuit takes place in the Porter-Thomas regime. We therefore  make the following ansatz for the output state $\rho_\cK$
\begin{align}
  \rho_\cK = \alpha \ket{\psi_d}\bra{\psi_d} + (1-\alpha) \frac \openone N\;\label{eq:ra}.
\end{align}
 As seen in Fig.~\ref{fig:pnp}, errors alter the shape of the Porter-Thomas distribution, approaching the uniform distribution as $\alpha \to 0$.

The cross entropy difference $\Delta \H$ defined in Eq.~\eqref{eq:5} is given by the probability distribution of $\log(p_U(x))$ where the bit-strings $x$ are sampled from the output $\rho_\cK$ of a circuit implementation with fidelity $\alpha$. Using Eq.~\eqref{eq:ra} and the Porter-Thomas distribution for $p_U(x)$ we obtain
\begin{align}
  \Pr_\alpha(z) = e^{z-e^z}\(1+\alpha\(e^z-1\)\)\;,\label{eq:pal}
\end{align}
where $z = \log(Np)$. If bit-strings are sampled uniformly, $-\log p_U(x)$ has a Gumbel distribution. We find a good fit between this expression and numerical simulations, see Fig.~\ref{fig:pal}. The value of $\alpha$ corresponding to a given Pauli error rate per gate can be estimated using Eq.~\eqref{eq:6}.

\section{Convergence to Porter-Thomas}\label{sec:pt}

\begin{figure}
  \centering\includegraphics[width=\columnwidth]{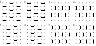}
  \caption{Layouts of CZ gates in a $6 \times 6$ qubit lattice. It is currently not possible to perform two CZ gates simultaneously in two neighboring superconducting qubits~\cite{barends_superconducting_2014,barends_digital_2015,kelly_state_2015,barends_digitized_2015}. We iterate over these arrangements sequentially, from 1 to 8.}
  \label{fig:two_dimers}
\end{figure}

In this section we report the results of numerical simulations on the required depth to approximate the Porter-Thomas distribution using planar quantum circuits that would be feasible to implement using state-of-the-art superconducting qubit platforms~\cite{barends_superconducting_2014,barends_digital_2015,kelly_state_2015,barends_digitized_2015}. The following circuits were chosen through numerical optimizations to minimize the convergence time to Porter-Thomas.
\begin{enumerate}
\item Start with a cycle of Hadamard gates (0 clock cycle).
\item Repeat for $d$ clock cycles:
  \begin{enumerate}
  \item Place controlled-phase (CZ) gates alternating between eight configurations similar to Fig.~\ref{fig:two_dimers}.  
   \item Place single-qubit gates chosen at random from
    the set $\{\X^{1/2}, \Y^{1/2}, \T\}$ at all qubits that are not occupied by
    the CZ gates at the same cycle (subject to the restrictions below). The gate $\X^{1/2}$ ($\Y^{1/2}$) is a $\pi/2$ rotation around the $X$ ($Y$) axis of the Bloch sphere, and the non-Clifford T gate is the diagonal matrix $\{0,e^{i \pi/4}\}$. 
  \end{enumerate}

\end{enumerate}
In addition, single-qubit gates are placed subject to the following rules: 
\begin{itemize}
\item Place a gate at qubit $q$ only if this qubit is occupied by a CZ
  gate in the previous cycle. 
\item Place a T gate at qubit $q$ if there
  are no single-qubit gates in the previous cycles at qubit $q$ except
  for the initial cycle of Hadamard gates. 
\item Any gate at qubit $q$ should
  be different from the gate at qubit $q$ in the previous cycle.
\end{itemize}

In the numerical study we calculate statistics corresponding to measurements in the computational (or Z) basis after each cycle. Because the CZ gates are diagonal in this basis, some gates before the measurement could be simplified away. The circuit would be harder to simplify if a cycle of Hadamards is applied before measuring in the Z basis. We did not apply a final cycle of Hadamards in the numerical study because it would double the computational run time, as the cycle of Hadamards would have to be undone after collecting statistics at cycle $t$ before moving to cycle $t+1$. We argue that the Porter-Thomas form of the output distribution, characteristic of chaotic systems, makes it unlikely that these circuits can be simplified substantially (see also Secs.~\ref{sec:intro} and~\ref{sec:ch}).

\begin{figure}
  \centering\includegraphics[width=\columnwidth]{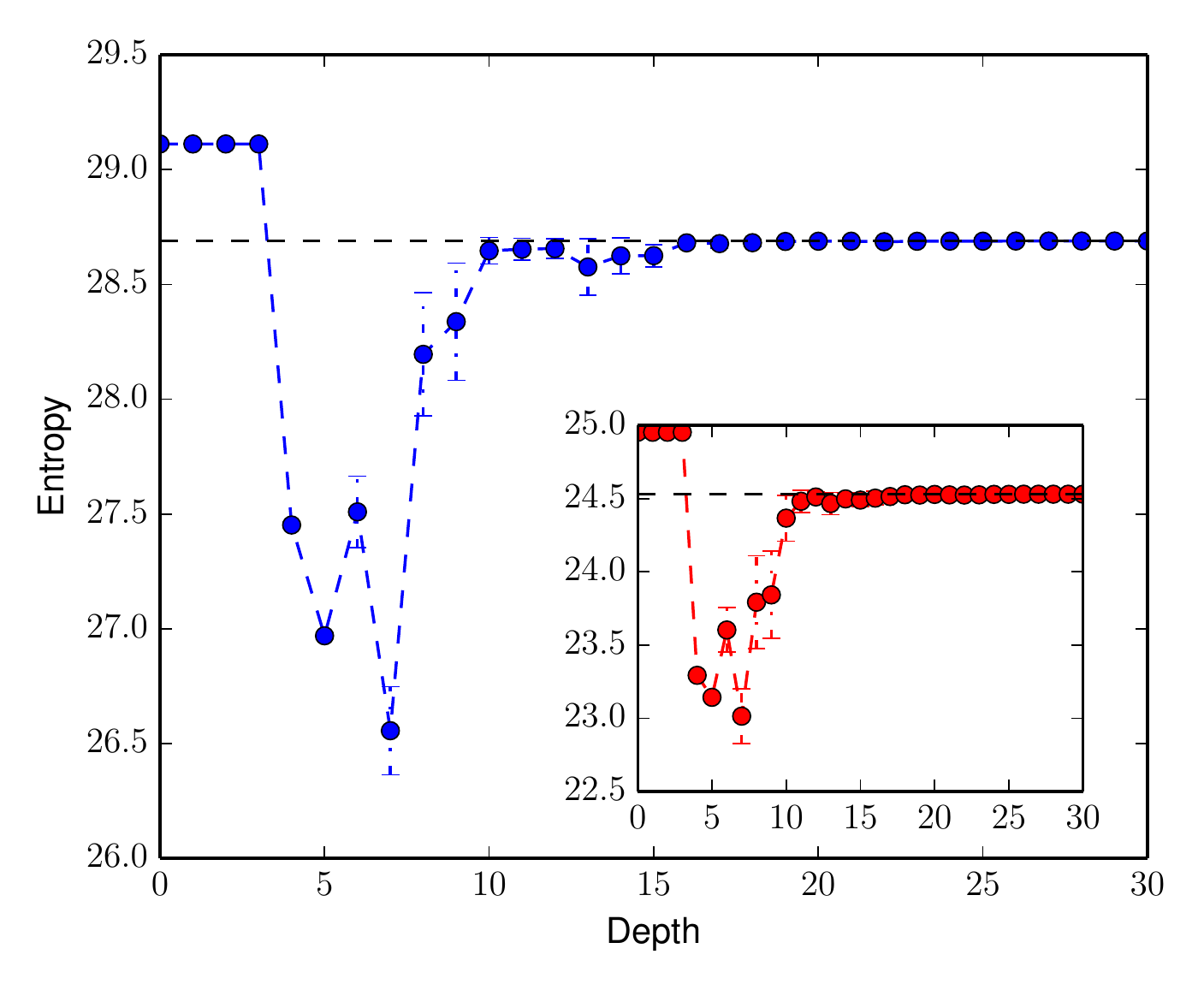}
  \caption{Mean entropy of the output distribution as a function of depth. The main figure pertains to circuits with $7 \times 6$ qubits, and the inset pertains to circuits with $6 \times 6$ qubits. The black dashed lines correspond to the entropy of the Porter-Thomas distribution.  Error bars are standard deviations among different circuit instances.
  }
  \label{fig:evd}
\end{figure}

Random circuits approximate a pseudo-random distribution~\cite{emerson2003pseudo,oliveira_efficient_2007} with logarithmic depth in a fully connected architecture~\cite{harrow2009random,weinstein_parameters_2008,brown_scrambling_2012}. These circuits can be embedded with depth proportional to $\sqrt{n}$, up to polylogarithmic factors in $n$, in a 2D lattice~\cite{beals2013efficient}. Consistent with our earlier discussion, we study how the entropy of the circuit output converges to the entropy of the Porter-Thomas distribution, Eq.~\eqref{eq:eu}. Figure~\ref{fig:sva} ($r=0$ line) shows that for all sizes of circuits up to $7 \times 6$ qubits, constructed according to the restrictions given above, our simulations reveal that the output distribution has the same entropy as the Porter-Thomas distribution. Figure~\ref{fig:evd} shows the output distribution entropy as a function of circuit depth. Circuits approach the Porter-Thomas regime with approximately ten cycles. Note that the initial entropy corresponds to the uniform distribution due to the first layer of Hadamards. Gates in the first cycles are diagonal and do not change the output entropy.

\begin{figure}
  \centering\includegraphics[width=\columnwidth]{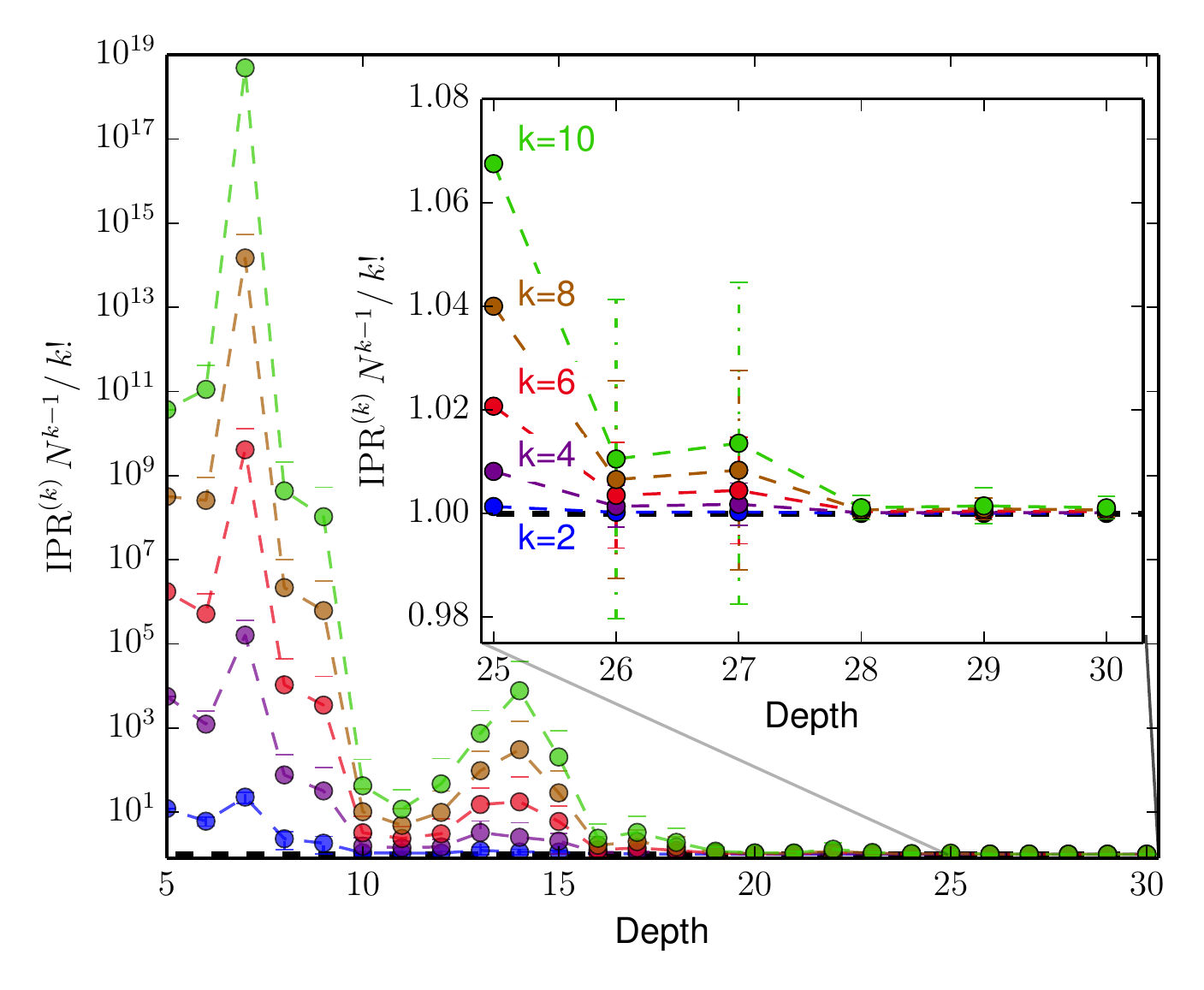}
  \caption{Mean normalized inverse participation ratios $k \in [2,..,10]$ of the output distribution ($\IPR^{(k)} \simeq N \avg{p^k}$) as a function of depth for circuits with $7 \times 6$ qubits. The black dashed line at the bottom corresponds to the Porter-Thomas distribution. Error bars correspond to the standard deviation between different circuit instances.}
  \label{fig:mvd}
\end{figure}

To develop intuition about the chaotic evolution of the wavefunction,
we focus on the degree of delocalization of the distribution $p_U(x_j)
$. The degree of delocalization is captured by the inverse participation
ratios $\IPR_t^{(k)} = \sum_j |\braket{x_j}
{\psi_t}|^{2k}$~\cite{brown_quantum_2008,de_luca_ergodicity_2013},
related to the moments of the distribution. If the wavefunction has
support over $\xi_t N$ local basis vectors, then $\IPR_t^{(k)} \propto
N^{-k+1} \xi_t^{-k}$.  As $t$ increases, $\xi_t \to 1$ and the
wavefunction becomes a pseudo-random vector sampled uniformly from
Hilbert space. At that point, finite moments of the distribution
converge to Porter-Thomas, $\IPR_t^{(k)} \to N^{-k+1}
k!$~\cite{harrow2009random,arnaud2008efficiency,brown2010convergence}. 
Importantly, we find numerically that convergence is achieved for
small order moments at a similar depth. This is evidenced in
Fig.~\ref{fig:mvd} for moments up to $k=10$ with circuits consisting
of $7 \times 6$ qubits.

\begin{figure}
  \centering\includegraphics[width=\columnwidth]{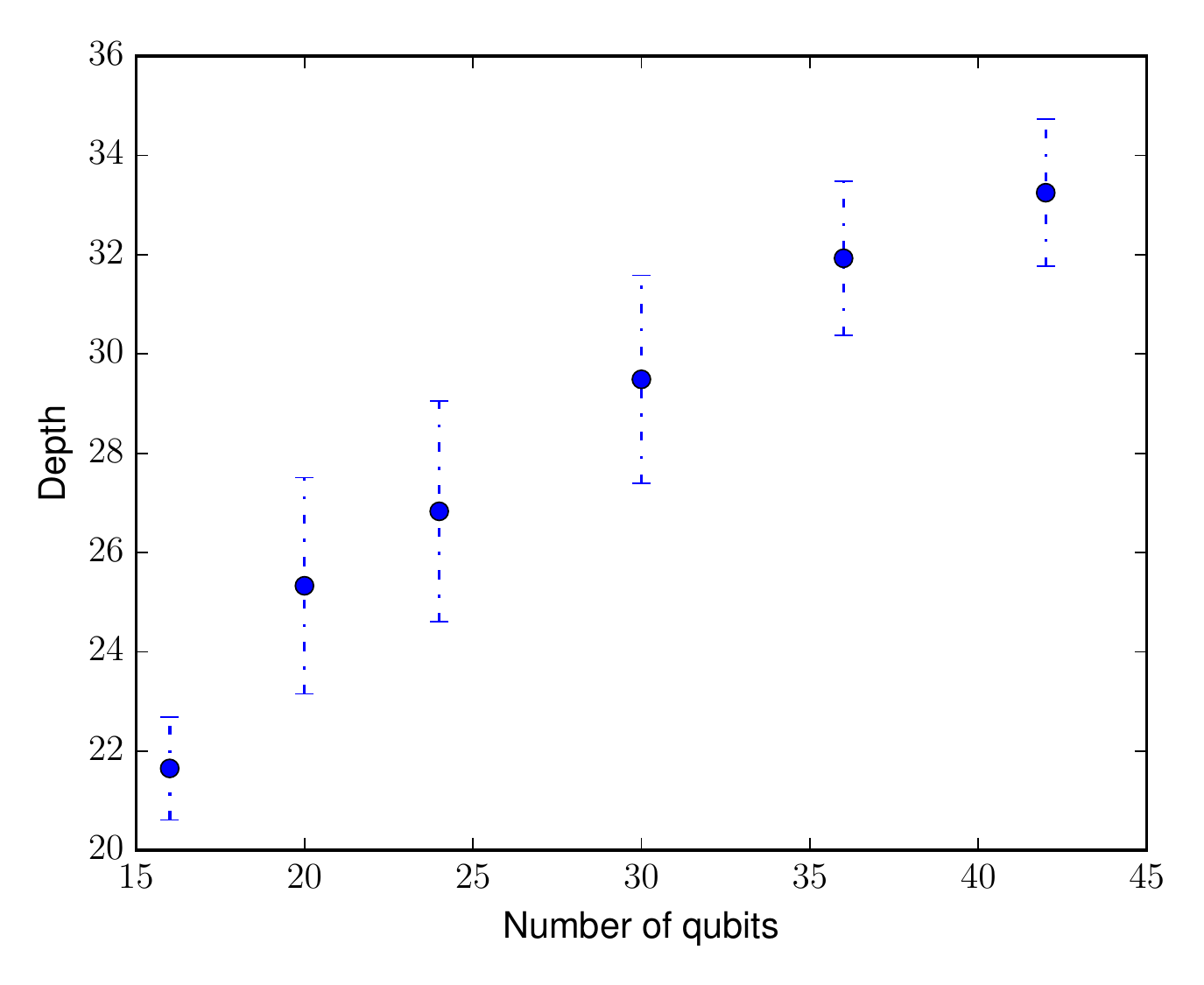}
  \caption{First cycle in a random circuit instance such that the entropy remains within $4$-sigma of the Porter-Thomas entropy during all the following cycles. Markers show the mean among instances and error bars correspond to the standard deviation among circuit instances. }
  \label{fig:pt_depth}
\end{figure}  

We also studied the expected convergence to Porter-Thomas with depth proportional to  $\sqrt{n}$ using a stronger criterion. The standard deviation of the entropy between different quantum states drawn from the Porter-Thomas distribution scales as $\approx 0.75\cdot 2^{-n/2}$. In Fig.~\ref{fig:pt_depth} we show the first cycle of each random circuit instance for which the entropy remains within $4$-sigma of the Porter-Thomas entropy during all the following cycles. These data indicates that the required depth to achieve this criteria grows sublinearly in $n$. We show a similar plot for circuits with denser layouts of $\CZ$ gates, which can be more appropriate for other qubit implementations, in App.~\ref{app:pt_nc}.

We note that a sublinear convergence to the second moment of the Porter-Thomas distribution is still faster than rigorously proven bounds for random circuits, such as Ref.~\cite{nakata2016efficient}. Interestingly, sparse IQP circuits achieve a similar property (so-called anticoncentration) with depth proportional to $\sqrt{n}$, up to polylogarithmic factors in $n$, in a 2D lattice~\cite{bremner16}. We have numerically verified that the output distribution of these circuits has the same entropy (up to small statistical fluctuations of order $2^{-n/2}$) as the Porter-Thomas distribution.

\section{Computational hardness of the classical sampling problem}\label{sec:ch}

The distribution $p_U(x) \propto 1/2^n$ is highly delocalized in the computational basis and in any basis obtained from local rotations of the computational basis. Therefore, it is impossible to estimate $p_U(x)$ for any $x$, even using a quantum computer, as doing so would require an exponential number of measurements. Nevertheless, the distribution $p_U(x)$ can be sampled efficiently by performing measurements on the state produced by the shallow random circuit $U$ on a quantum computer. In contrast, as we argued above from numerical simulations and the chaotic nature of the evolution, a classical algorithm can only sample from the distribution $p_U(x)$ if it can compute this function explicitly. This requires resources which grow exponentially in $n$, making the problem intractable even for modest sized random quantum circuits.

This intuitive argument can be made more rigorous in the asymptotic limit using computational complexity theory. Previous studies have introduced related sampling problems that a quantum computer can solve without having the ability to estimate $p_U(x)$~\cite{aaronson2003quantum,terhal2004adaptive,aaronson2005quantum,bremner_classical_2011,aaronson2011computational,aaronson2014equivalence,fujii2014impossibility,jozsa_classical_2014,bremner2015average,farhi_quantum_2016}. In this section we will extend the method used to show the computational hardness of sampling commuting random circuits (IQP)~\cite{bremner_classical_2011,bremner2015average} to the general case of universal random circuits. 

We will first describe the computational complexity class of estimating a probability $p_\cl(x)$ of a polynomial classical sampling algorithm. This is based on the fact that a random classical algorithm uses random bits, which is very different from the intrinsic randomness of quantum mechanics. We will then argue that approximating $p_U(x)$ belongs to a much harder complexity class, which implies that there does not exist an efficient classical sampling algorithm. 

A stronger recent conjecture states directly that no polynomial classical algorithm can estimate if $p_U(x)$ is above the median with bias better than $\sim 2^{-n}$~\cite{aaronson2016complexity}.

\subsection{General overview of the computational complexity argument }\label{sec:cca}

A classical sampling algorithm corresponds to the evaluation of a function
\begin{align}
  \label{eq:13}
  f(w,y) = x\;.
\end{align}
Here the bit-string $w = \{w_1,\ldots,w_k\}$ encodes the problem instance, $y$ is a vector of random bits $y = \{y_1,\ldots,y_\ell\}$ chosen uniformly and $x$ is the output bit-string. For fixed $w$ and $x$, the number $W_x$ of solution vectors $y$ of Eq.~\eqref{eq:13}  defines the probability $q(x) =  W_x/2^\ell$ of getting a sample $x$. Assume that evaluating the function $f$ can be done in a time which scales polynomially in the number of input bits $k+\ell$, with $\ell$ polynomial in $k$. Then, the problem of determining if there is a solution vector $y$ to Eq.~\eqref{eq:13} with fixed $w$ and $x$ belongs to the complexity class NP. A complexity theory abstraction that solves this general problem is called an {\it NP-oracle}. An important result in computer science, the so-called Stockmeyer Counting Theorem~\cite{stockmeyer1983complexity}, states that probabilistically approximating the number of solutions $W_x$, and therefore $q(x)$, to within a multiplicative factor, can also be performed with an NP oracle, see App.~\ref{app:stockmeyer}. 

A classical sampling algorithm simulating a quantum random circuit $U$ must output bit-strings $x$ with probability $q(x)$ approximating $p_U(x)$. The input vector $w$ to the corresponding function $f(w,y)$ is a description of the circuit $U$, which is polynomial in the number of qubits $n$. It has been shown that,  in the case of commuting quantum circuits, the function $p_U(x) = |\braket{x}{\psi}|^2$ encodes the partition function of a random complex Ising model~\cite{bremner_classical_2011,bremner2015average} 
\begin{align}
  \label{eq:pis}
  \braket{x}{\psi} = \lambda \sum_{ s} e^{ i \theta H_x(s)} \;, \quad H_x(s) = h_x\!\cdot\! s+ s\!\cdot \!\hat J \!\cdot\! s\;,
\end{align}
where $H_x(s)$ is a classical energy, $ s$ is a vector of classical spins $\pm 1$, $ h_x$ is a vector of local fields, $\hat J$ is the coupling matrix, $i \theta$ is the inverse imaginary temperature and $\lambda$ is a scaling constant. The partition function can also be written as $\sum_j M_j e^{i \theta E_j}$ where $M_j$ is the number of solutions $s$ to the equation $H_x(s) = E_j$. In general, the $M_j$'s grow exponentially in the number of classical spins. 

The partition function at low {\it real-valued} temperatures $\T$ (with $\theta=i/T$) is hard to approximate only because the sum in Eq.~\eqref{eq:pis} is dominated by low energy states. The Stockmeyer Counting Theorem implies that probabilistically approximating the corresponding $M_j$ within a multiplicative error can be done with an NP-oracle, because for any given $s$ the energy $H_x(s)$ can be calculated efficiently. This results in a multiplicative error estimation of the partition function. In contrast, for purely imaginary temperatures $i/\theta$, the sum $\sum_j M_j e^{i \theta E_j}$ is determined by the intricate cancellations between individual terms, each exponentially large in magnitude. A discussion of this cancellation for the case of random circuits is given in the next subsection. An approximation of $M_j$ with multiplicative error is not sufficient to estimate the partition function. Therefore, the case with purely imaginary temperatures is much harder than the real-valued case. 

These intuitive arguments are supported by the strongly held conjecture in computational complexity theory that probabilistically approximating partition functions with purely imaginary temperatures is much harder, in the worst case, than any problem which can be solved NP oracle~\cite{bremner_classical_2011,fujii2013quantum,goldberg_complexity_2014}. Reference~\cite{bremner2015average} argues that because random instances of Ising models have no structure making them easier, the same conjecture applies to any sufficiently large fraction of partition functions of random complex Ising models.

Assume now that there exists an approximate classical sampling algorithm for the distribution $p_U$ with asymptotic complexity polynomial in $n$ and small distance in the $\ell_1$ norm. From the convergence of the second moment of $p_U$ to the Porter-Thomas distribution found numerically, it would then follow from the proof in Ref.~\cite{bremner2015average} that a fraction of these probabilities could be probabilistically approximated  with multiplicative error using an NP-oracle, see App.~\ref{sec:maz}. As argued above, this is implausible for a complex partition function with the general form of Eq.~\eqref{eq:pis}. We will show in the next section that $p_U(x)$ can be mapped directly to the partition function of a quasi three-dimensional random Ising model, with no apparent structure that makes it easier to approximate than a random instance. If we conjecture that a sufficient large fraction of these instances is as hard to approximate as the worst case, we must conclude that such an efficient classical sampling cannot be achieved.

\subsection{The partition function for random circuits}\label{sec:pf}
 While our approach for mapping circuits to partition functions can be applied to any circuit, we focus here on the particular case of a quantum circuit $U$ as described in  Sec.~\ref{sec:pt}. Known algorithms for mapping universal quantum circuits to partition functions of complex Ising models use polynomial reductions to a universal gate set~\cite{lidar_quantum_2004,geraci_classical_2010,de2011quantum}. Here we provide a direct construction, which allows us to define a random ensemble of Ising models without apparent structure. We represent the circuit by a product of unitary matrices $U^{(t)}$ corresponding to different clock cycles $t$, with the $0$-th cycle formed by Hadamard gates. We introduce the following notation for the amplitude of a particular bit-string after the final cycle of the circuit, 
\begin{equation}
 \langle x|\psi_d\rangle=\sum_{\{\sigma^{t}\}}\prod_{t=0}^{d}\bra{\sigma^{t}}U^{(t)}\ket{\sigma^{t-1}} ,\quad \ket{\sigma^{d}}=\ket{x}.\label{eq:path}
\end{equation}
Here $\ket{ \sigma^t}=\otimes_{j=1}^{n}\ket{\sigma^t_j}$ and the  assignments $\sigma_j^t=\pm 1$ correspond to the states $\ket{ 0}$ and $\ket{1}$ of the  $j$-th qubit, respectively.  The expression (\ref{eq:path}) can be viewed as a Feynman path integral with individual paths $\{\sigma^{-1},\sigma^{0},\ldots,\sigma^{d}\}$
formed by a sequence of the computational basis states of the $n$-qubit system. The  initial condition for each path  corresponds to $\sigma^{-1}_j=0$ for all qubits and the final point corresponds to  
$\ket{\sigma^{d}}=\ket{x}$. 

Assuming that a  $\T$ gate is applied to qubit $j$ at the cycle $t$, the indices of the matrix $\bra{\sigma^{t}}U^{(t)}\ket{\sigma^{t-1}}$ will be equal to each other, i.e. $\sigma_j^{t}=\sigma^{t-1}_j$. 
A similar property applies to the $\CZ$  gate as well. The state  of a qubit can only flip under the action of the gates $\H$, $\X^{1/2}$ or $\Y^{1/2}$.
We refer to these as two-sparse gates as they contain two nonzero elements in each row and column (unlike $\T$ and $\CZ$). This observation allows us to rewrite the path integral representation in a more economic fashion.

Through the circuit, each qubit $j$ has a sequence of two-sparse gates applied to it. We denote the length of this sequence as $d(j)+1$ (this includes the $0$-th cycle formed by a layer  of Hadamard gates applied to each qubit). 
In a given  path  the qubit $j$ goes through the  sequence of spin states $\{s_j^k\}_{k=0}^{d(j)},$  where,  as before, we have $s_j^k=\pm 1$.  The value of  $s_j^k$ in the sequence determines the state of the qubit  {\it immediately after}  the action of the $k$-th two-sparse gate. The last element in the sequence is fixed   
by the assignment of bits in the bit-string $x$,
\begin{equation}
s_{j}^{d(j)}=x^{(j)}\;,\quad j\in [1\isep n]\;.\label{eq:bc}
\end{equation}
Therefore, an individual path in the path integral can  be encoded by the set of $G=\sum_{j=1}^{n}d(j)$ binary variables $s=\{s_j^k\}$ with  $j\in[1\isep n]$ and  $k\in[0 \isep d(j)-1]$. One can easily see from  the  explicit form of the two-sparse gates that  the absolute values of the probability amplitudes associated  with different  paths are all the same and equal to   $2^{-G/2}$. Using this fact we write the  path integral (\ref{eq:path}) in the following form 
\begin{equation}
\langle x| \psi_d\rangle=2^{-G/2} \sum_{s}\exp\left(\frac{i\pi}{4} H_s(x) \right)\;.\label{eq:path1} 
\end{equation}
Here $\exp(i\pi H_s(x)/4)$ is  a phase factor associated with each path that depends explicitly on the end-point condition (\ref{eq:bc}).

The value of  the phase  $\pi H_s/4$ is accumulated as a sum of discrete phase changes that are associated with individual gates. For the $k$-th two-sparse gate applied to qubit $j$ we introduce the coefficient $\alpha_j^k$ such that  $\alpha_j^k=1$ if the gate is  $\X^{1/2}$ and  $\alpha_j^k=0$ if the gate is $\Y^{1/2}$. 
 Thus, the  total phase change accumulated from the application of $\X^{1/2}$  and $\Y^{1/2}$  gates equals
\begin{align}
\hspace{-0.1in} \frac{i\pi}{4} H_{s}^{\X^{1/2}}(x) &=\frac{i \pi}{2} \sum_{j=1}^{n}\sum_{k=0}^{d(j)}\alpha_j^k \frac{1+s_{j}^{k-1}s_{j}^{k}}{2}\;, \label{eq:phX} \\
\frac{i\pi}{4} H_{s}^{\Y^{1/2}}(x) &=i \pi  \sum_{j=1}^{n}\sum_{k=0}^{d(j)}(1-\alpha_j^k) \frac{1-s_{j}^{k-1}}{2}\frac{1+s_j^k}{2}\;\nonumber\;.
\end{align}
As mentioned above, the dependence on $x$ arises due to the boundary condition (\ref{eq:bc}).  Note that we have omitted constant phase terms that do not depend on the path $s$.

We now describe the phase change from the action of  gates $\T$ and $\CZ$.  We introduce  coefficients $d(j,t)$  equal to the number of two-sparse  gates applied to qubit $j$ over the first $t$ cycles (including the $0$-th cycle  of Hadamard gates).  We also introduce coefficients $\tau_j^t$ such that $\tau_j^t=1$ if a $\T$ gate is applied at cycle $t$ to qubit $j$ and $\tau_j^t=0$ otherwise.
Then the total  phase accumulated  from the action of the $\T$ gates equals
\begin{align}
\frac{i\pi}{4} H_{s}^{T}(x)=\frac{i \pi}{4} \sum_{j=1}^{n}\sum_{t=0}^{d}\tau_j^t \frac{1-s_{j}^{d(j,t)}}{2}\label{eq:phT}\;.
\end{align}
For a given pair of qubits $(i,j)$, we introduce coefficients $z_{ij}^{t}$ such that   $z_{ij}^{t}=1$ if  a $\CZ$ gate is applied to the  qubit pair during cycle $t$ and  $z_{ij}^{t}=0$ otherwise.
The total  phase accumulated  from the action of the $\CZ$ gates equals
\begin{multline}
\frac{ i\pi}{4} H_{s}^{\CZ}(x)\\=i \pi \sum_{i=1}^{n}\sum_{j=1}^{i-1}\sum_{t=0}^{d}z_{ij}^t  \frac{ 1-s_{i}^{d(i,t)}}{2}\frac{1-s_{j}^{d(j,t)}}{2}.\label{eq:phCZ}
\end{multline}

One can see from comparing   (\ref{eq:path1}) with (\ref{eq:phX})-(\ref{eq:phCZ}) that the wavefunction amplitudes $ \langle x|\psi_d\rangle$ take the form of a partition function of a classical Ising model 
with energy $H_s$ for a state $s$ and purely imaginary inverse temperature $i \pi/4$. The total phase for each path  takes 8 distinct values (mod 2$\pi$)  equal to $[0,\pi/4 \isep 7\pi/4]$. 
The function  $H_s(x)$  can be written as a sum of three different types of terms 
\begin{equation}
H_s(x)=H_{s}^{(0)}+H_{s}^{(1)}+H^{(2)}\;.\label{eq:Hsx}
\end{equation}
Here
\begin{multline}
H_{s}^{(0)}=\sum_{i=1}^{n}\sum_{k=1}^{d(i)-1}h_i s_i \\+\sum_{i=1}^{n}\sum_{j=1}^{i-1}\sum_{k=1}^{d(i)-1}\sum_{l=1}^{d(j)-1}{\cal J}_{ij}^{kl}s_i^k s_j^l.\label{eq:Hs}
\end{multline}
is  the  energy term quadratic in spin variables and  expressed in terms of the Ising coupling coefficients ${\cal J}_{ij}^{kl}$ and local fields $h_i^k$ to be given below. It does not depend on the spin configuration $x$ of  the final point on the paths. $H_{s}^{(1)}$ is a bilinear function of  Ising spin  variables $s$ and $x$ 
\begin{equation}
H_{s}^{(1)}(x)=\sum_{i=1}^{n}\sum_{j=1}^{n}\sum_{k=1}^{d(i)-1}b_{ij}^{k} s_i^k x^{(j)}\;.\label{eq:H1}
\end{equation}
 The term  $H^{(2)}(x)$ depends on $x$ but not  $s$. For brevity, we do not provide its explicit form.

The local fields  $h_j$ are  computed as
\begin{equation}
h_i^k=\alpha_i^{k+1}-\alpha_i^k-\frac{1}{2}J _{i}^{k}-\sum_{j=1}^{n}\sum_{l=1}^{d(j)} J^{k\,l}_{ij}
\end{equation}
and 
 the coupling constants  ${\cal J}_{ij}^{kl}$ equal
\begin{equation}
{\cal J}_{ij}^{kl}=J_{ij}^{kl}+\frac{1}{2}\delta_{i,j}(\delta_{k-1,l}+\delta_{k,l-1})\left(2 \alpha_{i}^{(k+l+1)/2}-1\right)\label{eq:cJ}
\end{equation}
where
\begin{equation}
J_{ij}^{kl}=\sum_{t=1}^{d}\delta_{k,d(i,t)}\delta_{l,d(j,t)} z_{ij}^{t}\;,\label{eq:JJ}
\end{equation}
and
\begin{equation}
J_i^k=\sum_{t=1}^{d}\delta_{k,d(i,t)}\tau_i^t\;.\label{eq:J}
\end{equation}
The coupling coefficients $b_{ij}^{k}$ in (\ref{eq:H1}) equal
\begin{equation}
b_{ij}^{k}=\delta_{k,d(i)-1}\delta_{ij}(2\alpha_{j}^{d(j)}-1)+J_{ij}^{k d(j)}\;.
\end{equation}
  The Ising coupling for spin $s_{j}^{d(j)}=x^{(j)}$ induces an additional local field   $\sum_{j=1}^{n}\sum_{k=1}^{d(i)-1} b_{ij}^{k} x^{(j)} $ on spin  $s_{i}^{k} $ as shown in (\ref{eq:Hs}).
 
  To understand the structure of the  graph defined by the Ising couplings  (\ref{eq:cJ})  we study the statistical ensemble of ${\cal J}_{ij}^{kl}$. For simplicity, we will analyze circuits composed of $d$ layers, each layer consisting of a cycle of single-qubit gates followed by a cycle of two-qubit CZ gates (see App.~\ref{app:pt_nc}). We also assume here that the layout of the two-qubit CZ gates is random, and that in the single-qubit gate cycles the gates $X^{1/2}$,  $Y^{1/2}$, and  $T$ are applied to a qubit with equal  probabilities. 

To describe the  evolution of qubit states under the action of the  gates we need to introduce a third dimension to describe the graph of  the Ising couplings, Eq.~\eqref{eq:cJ}.
For each qubit $j$ we introduce a ``worldline'' with a grid of points  enumerated by $t\in [1 \isep d]$, each corresponding to a layer.
We denote the layer numbers where the function $d(j,t)$ increases from $k-1$ to $k$ by a two-sparse gate applied to qubit $j$ as $t_j^k$. We associate Ising spins $\{s_j^k\}_{k=0}^{d(j)-1}$ to vertices   of the graph  located at the grid points $\{t_j^k\}$  along the   worldline $j$.

 Consider a pair of vertices corresponding  to spins $s_i^k$ and $s_j^l$ associated with the two  adjacent qubits $i$ and $j$. Then the   coefficient $J_{ij}^{kl}$  equals to the number of applied CZ gates that couple qubits $i$ an $j$ during the sequence of layers $[\max(t_i^k,t_j^l) \isep (\min(t_i^{k+1}, t_j^{l+1})-1)]$.
 The distribution of $J_{ij}^{kl}$ can be written in the following form 
\begin{equation}
{\rm Pr}[J_{ij}^{kl}=r]\equiv P(r)= \sum_{q=0}^{\infty} p(r|q) p(q)\,,\label{eq:PrJ}
\end{equation}
Here 
$p(q)=\frac{8}{9}\left(\frac{1}{3}\right)^{2q}$ 
is the probability of having  {\it no}  two-sparse gates applied to   qubits  $i$ and   $j$  for $q$ layers and then having   a  two-sparse gate applied to at least one  of them in the $(q+1)^{\rm st}$ layer. Also
$p(r|q) =\binom{q+1}{r} p_{\rm CZ}^{r} (1-p_{\rm CZ})^{q+1-r}$     
  is the probability of having $r$  CZ gates over $q+1$ layers applied between a given pair of neighboring qubits.  Finally, we have for $P(r)$
\begin{equation}
P(r)=\begin{dcases} 
\frac{1-p_{\rm CZ}}{1+p_{\rm CZ}/8},  & r=0   \\
\frac{9}{1+p_{\rm CZ}/8}\,\left(\frac{p_{\rm CZ}/8}{1+p_{\rm CZ}/8}\right)^r & r>0 \;.
\end{dcases}\label{eq:Prex}
\end{equation}
For a square grid of qubits  $p_{\rm CZ}\simeq 1/4$. One can see from (\ref{eq:Prex}) that for $r\geq 1$ the distribution ${\rm Pr}[J_{ij}^{kl}=r]$ decays exponentially with $r$ and $P(r+1)/P(r)\simeq p_{\rm CZ}/8\simeq 1/32$. Therefore, the most likely values of   $J_{ij}^{kl}$  are 0, corresponding to 
 the probability  $P(0)\simeq\,1-p_{\rm CZ}$, and $1$, corresponding to the probability $P(1)\simeq 9p_{\rm CZ} /8$. The high probability of having no traversal couplings between qubits relates to the comparatively slow growth of the treewidth, see App.~\ref{app:treewidth}.
   
For fixed qubit indexes $(i,j)$, it is of interest to derive the conditional distribution $\mathfrak{p}(l|k)$ for spin $s_i^k$ to couple to spin $s_j^l$.
To obtain it we first introduce the probability $\mathfrak{p}_k(t)$ corresponding to the condition  $t_i^k=t$
 of having  the $k$-th  vertex located exactly at the layer $t$ of a given worldline. Not too close to the end of the  circuit  ($d-t\gg \sqrt{d}$) we have 
  \begin{equation}
\mathfrak{p}_k(t)=\binom{t-1}{k-1}\left( \frac{1}{3}\right)^{t-k}\left(\frac{2}{3}\right)^k,\quad \sum_{t=k}^{\infty}\mathfrak{p}_k(t)=1,\label{eq:pkt}
\end{equation}
Similarly, the probability $\mathfrak{p}^t(l)$ of having  exactly $l$ vertices located  within $t$ layers of a given worldline ($t_j^l\leq t$) equals
\begin{equation}
\mathfrak{p}^t(l)=\binom{t}{l}\left( \frac{1}{3}\right)^{t-l}\left(\frac{2}{3}\right)^l,\quad \sum_{l=0}^{t}\mathfrak{p}^t(l)=1\;.\label{eq:ptk}
\end{equation}
The above  conditional distribution $\mathfrak{p}(l|k)$ of the values of $l$ given $k$ equals
\begin{equation}
\mathfrak{p}(l|k)=\sum_{t}\mathfrak{p}^t(l)\mathfrak{p}_k(t)\;.
\end{equation}
Approximating  the binomial coefficients with the Stirling formula we obtain 
\begin{equation}
\mathfrak{p}(l|k)\simeq \sqrt{ \frac{3}{2\pi(k+l)}}\,\exp\left(-\frac{3(k-l)^2}{2(k+l)}\right)\;.\label{eq:Gauss}
\end{equation}
The above equation is asymptotically correct for $k,l$ not to close to the start and end points of the circuit, and $|k-l|\ll d$.

In summary, the coupling graph corresponding to the coefficients ${\mathcal J}_{ij}^{kl}$ represents a quasi three-dimensional structure formed by worldline corresponding to qubits located  on a 2D lattice. According to (\ref{eq:cJ}),  in the same worldline only neighboring vertices are  coupled. The strength of the coupling is $\pm 1/2$ depending on the type of the two-sparse gate. In general,   each vertex can be ``laterally" coupled to other vertices located on the neighboring  worldlines. The probability distribution  of the coupling coefficients has exponential form,  Eq.~\eqref{eq:Prex}. Differences between the vertex  indices that are involved in the lateral couplings obey a local  Gaussian distribution, Eq.~\eqref{eq:Gauss}.

Finally, note that Eq.~\eqref{eq:path1} can be written in the form  $\braket x {\psi_d} = 2^{-G/2} Z$, where $Z = \sum_{j=0}^7 M_j e^{i \frac {2\pi} 8  E_j}$ is a partition function, the $E_j$'s are different energies of the Ising model (mod 8) and $M_j \sim 2^{G}$. Furthermore, for a delocalized state $|\braket x \psi| \sim 2^{-n/2} $. Therefore, the partition function $|Z| \sim 2^{(G-n)/2}$ is exponentially smaller in $G$ than the individual terms $M_j$ in its sum. This very strong cancellation prevents any efficient algorithm from being able to accurately estimate the quantity $\braket x \psi$ (see also App.~\ref{app:feynman}).

Note that  if a quantum circuit uses only Clifford gates (not $\T$ gates), the total phase for each spin configuration in the partition function (mod 2$\pi$) is restricted to $[0,\pi/2,\pi,3\pi/2]$. In these case,  the corresponding partition function can be calculated efficiently~\cite{gottesman1998heisenberg,fujii2013quantum,goldberg_complexity_2014}.

\section{Conclusion}
In the near future, quantum computers without error correction will be able to approximately sample the output of random quantum circuits which state-of-the-art classical computers cannot simulate~\cite{aaronson2003quantum,terhal2004adaptive,aaronson2005quantum,bremner_classical_2011,aaronson2011computational,fujii2013quantum,aaronson2014equivalence,fujii2014impossibility,jozsa_classical_2014,bremner2015average,farhi_quantum_2016}. We have introduced a well-defined metric for this computational task. If an experimental quantum device achieves a cross entropy difference surpassing the performance of the state-of-the-art classical competition, this will be a first demonstration of quantum supremacy~\cite{preskill_2012}. The cross entropy can be measured up to the quantum supremacy frontier with the help of supercomputers. After that point it can be extrapolated by varying the number of qubits, the number of non Clifford gates~\cite{bravyi_improved_2016}, and/or the circuit depth~\cite{markov_simulating_2008,aaronson2016complexity}. Furthermore, the cross entropy can be approximated independently from estimates of the circuit fidelity. Quantum supremacy can be claimed if the theoretical estimates are in good agreement with the experimental extrapolations.

A crucial aspect of a near-term quantum supremacy proposal is that the computational task can only be performed classically through a direct simulation with cost exponential in the number of qubits. Direct simulations are required for chaotic systems, such as random quantum circuits~\cite{schack_hypersensitivity_1993,emerson2003pseudo,scott_hypersensitivity_2006}. A simulation can be done in several ways: evolving the full wavefunction; calculating matrix elements of the circuit unitary with tensor contractions~\cite{markov_simulating_2008,aaronson2016complexity}; using the stabilizer formalism~\cite{bravyi_improved_2016}; or summing a significant fraction of the corresponding Feynman paths in the partition function of an Ising model with imaginary temperature, see App.~\ref{app:feynman}. We study the cost of all these algorithms and conclude that, with state-of-the-art supercomputers, they fail for universal random circuits with more than approximately 48 qubits and depth $\sim 40$. 

We related the computational hardness of this problem, originating from the chaotic evolution of the wavefunction, to the sign problem emerging from the cancellation of exponentially large terms in a partition function of an Ising model with imaginary temperature. This finding is made more rigorous by results in computational complexity theory~\cite{bremner_classical_2011,fujii2013quantum,goldberg_complexity_2014,bremner2015average}. Following previous works~\cite{aaronson2011computational,kalai2014gaussian,arkhipov2015bosonsampling,leverrier2015analysis,bremner2015average}, we argue that, under certain assumptions, there does not exist an efficient classical algorithm which can sample the output of a random quantum circuit with a constant error (in the $\ell_1$ norm) in the limit of a large number of qubits $n$ (see Eq.~\eqref{eq:cl1}). Unfortunately, achieving a constant error in the limit of large $n$ requires a fault tolerant quantum computer, which will not be available in the near term~\cite{kalai2014gaussian,arkhipov2015bosonsampling,rahimi2016sufficient}. Nonetheless, it has been argued, also using computational complexity theory, that the \emph{exact} output distribution of certain quantum circuits with a constant probability of error per gate is  also asymptotically hard to simulate classically~\cite{fujii2014computational}.

A specific figure of merit for a well defined computational task, naturally related to fidelity, as well as an accurate error model, are equally crucial for establishing quantum supremacy in the near-term. This is absent from previous experimental results with quantum systems which can not be simulated directly~\cite{trotzky2012probing,lanting_entanglement_2014,boixo_evidence_2014,boixo_computational_2014,albash2015reexamining,isakov_understanding_2015,jiang_scaling_2016}. Without this, it is not clear if divergences between the experimental data and classical numerical methods~\cite{trotzky2012probing,albash2015reexamining} are due to the effect of noise or other unaccounted sources. Furthermore, we note that the numerical simulation and experimental curves in Ref.~\cite{trotzky2012probing} are reasonably well fitted by a rescaled cosine. Therefore, these curves can be approximately extrapolated efficiently classically. 

Finally, the problem of sampling from the output distribution defined by a random quantum circuit is a general, well known, computational task. A device which qualitatively outperforms state-of-the-art classical computers in this task is clearly not simply a device `simulating itself'. 

The evaluation of effective error models for large scale universal quantum circuits is a difficult theoretical and experimental problem  due to their complex nature. Therefore, existing proposals involve an expensive additional unitary transformation to the initial state~\cite{emerson_scalable_2005} or are restricted to non-universal circuits~\cite{flammia2011direct}. 
Our proposal based on experimental measurements of  the cross entropy, represents a novel way of characterizing and validating digital error models, and open quantum system theory in general.  The method introduced here can also be applied to other systems, such as continuous chaotic Hamiltonian evolutions.

\begin{acknowledgments}
  \label{sec: Acknowledgment}
  We specially acknowledge Mikhail Smelyanskiy, from the Parallel Computing Lab, Intel Corporation, who performed the simulations of circuits with $6 \times 6$ and $7 \times 6$ qubits and wrote Appendix~\ref{app:qsd}. 
  We would like to acknowledge Ashley Montanaro for multiple suggestions, specially regarding Sec.~\ref{sec:ch}.
  We would like to thank Scott Aaronson, Austin Fowler, Igor Markov,
  Masoud Mohseni and Eleanor Rieffel for discussions.
  The authors also thank Jeff Hammond, from the Parallel Computing Lab, Intel Corporation, for his useful insights into MPI run-time performance and
  scalability.
  This research used resources of the National Energy
  Research Scientific Computing Center, a DOE Office of Science User
  Facility supported by the Office of Science of the U.S. Department
  of Energy under Contract No. DEAC02-05CH11231. MJB has received financial support from the Australian Research Council via the Future Fellowship scheme (Project No. FT110101044).
\end{acknowledgments}

\appendix

\section{Residual correlations after discrete errors}\label{app:correlations}

\begin{figure}
  \centering
  \includegraphics[width=\columnwidth]{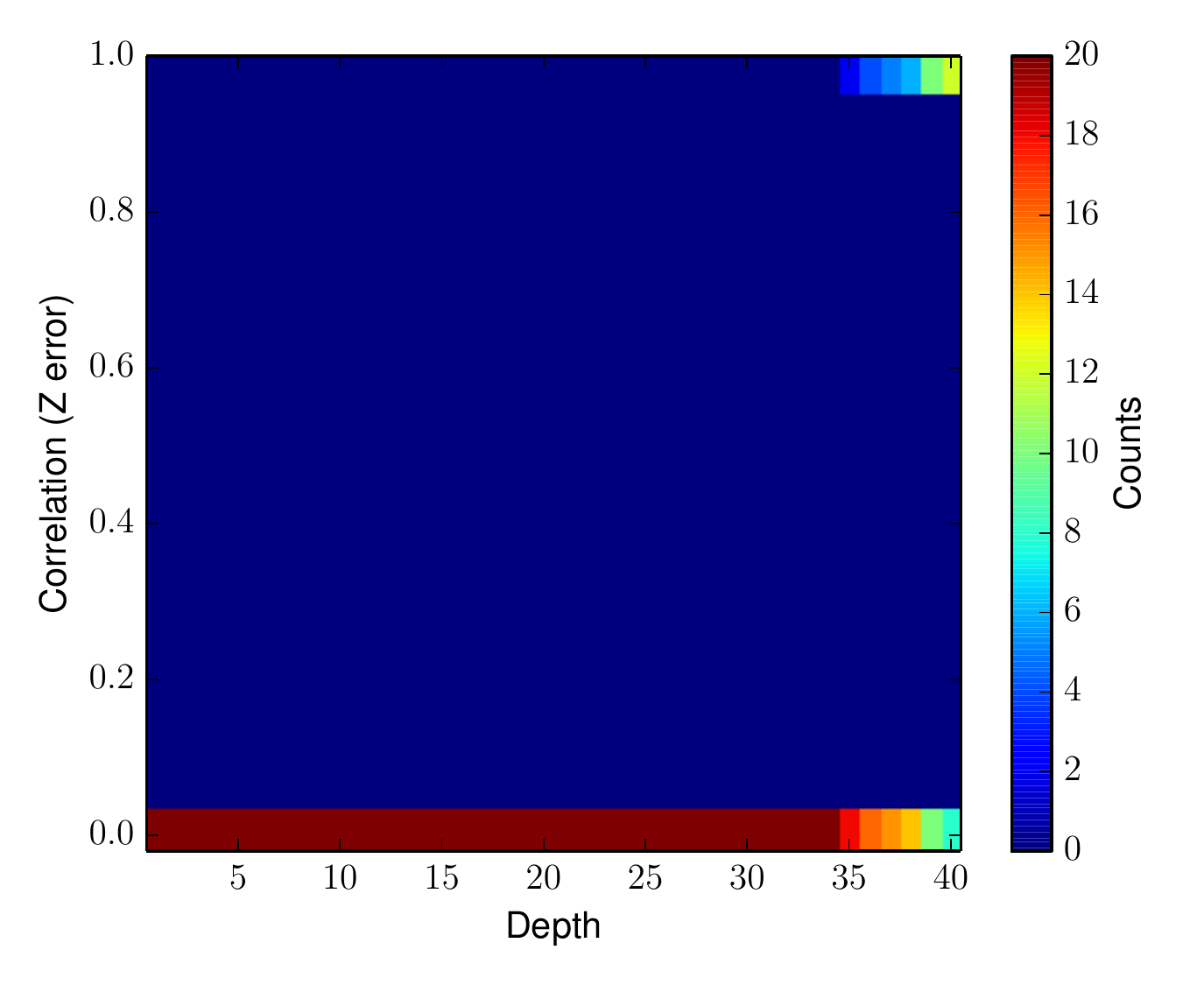}
  \caption{Two-dimensional histogram of residual correlations after a single $\Z$ error (phase-flip) is applied at different depths. We calculate numerically the correlation between the output of the circuit of Fig.~\ref{fig:1error}, with $5 \times 4$ qubits and total depth 40, and the output  when a phase flip is applied to one of the $20$ qubits.  }
  \label{fig:zcorr}
\end{figure}

\begin{figure}
  \centering
  \includegraphics[width=\columnwidth]{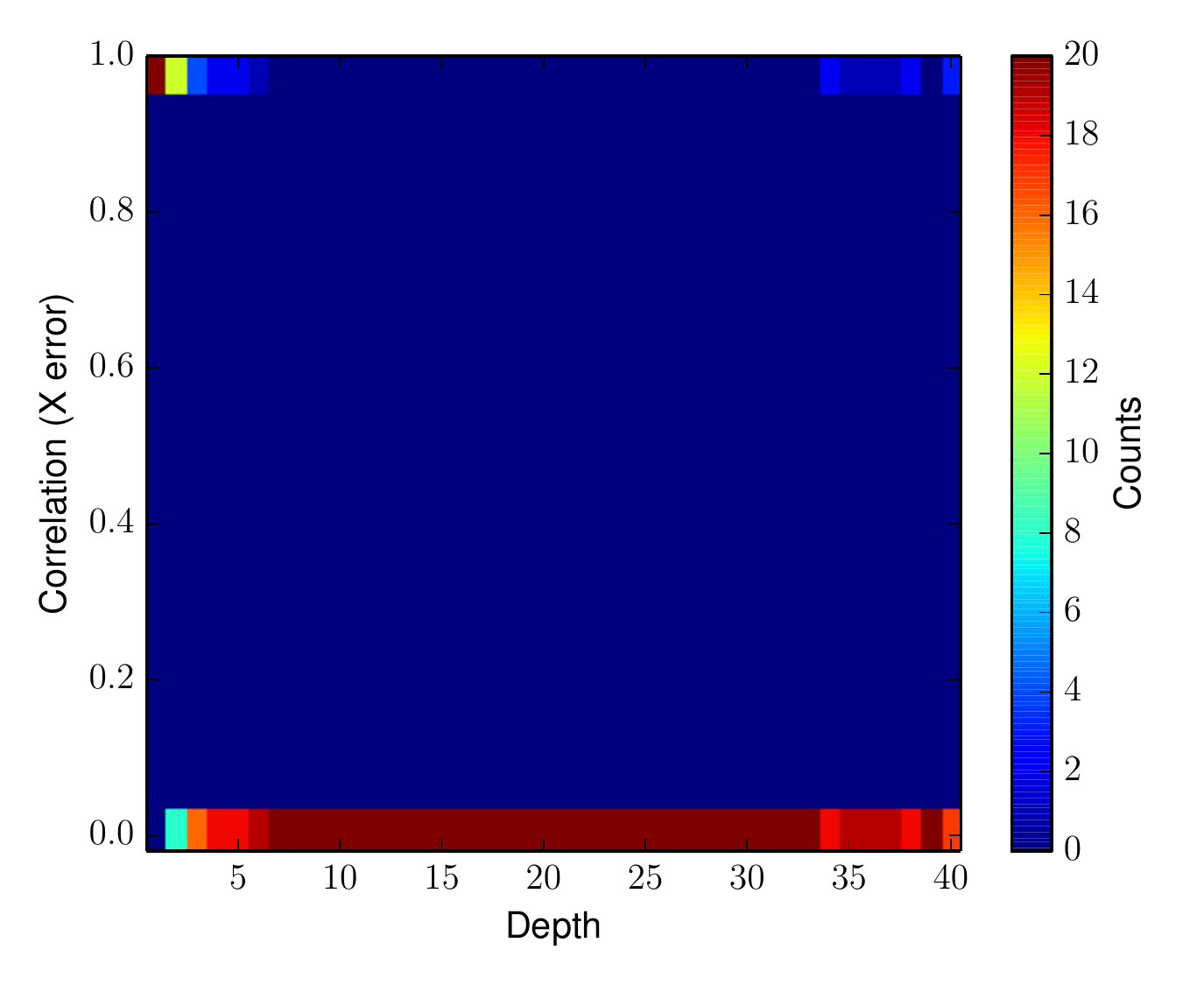}
  \caption{Two-dimensional histogram of residual correlations for a single $\X$ error (bit-flip) applied at different depths. Same circuit as in  Fig.~\ref{fig:1error} and Fig.~\ref{fig:xcorr}.}
  \label{fig:xcorr}
\end{figure}

In this appendix we analyze numerically the residual correlations between the output of an ideal circuit and the output when a single $\X$ error (bit-flip) or $\Z$ error (phase-flip) is applied to one of the qubits. This residual correlation is responsible for the slight upper curvature seen in the red line in Fig.~\ref{fig:1error}. It is also principally responsible for the small disparity between the cross entropy difference and the estimated fidelity seen in Fig.~\ref{fig:sva}. 

Figure.~\ref{fig:zcorr} shows the residual correlation for a single $\Z$ error (phase-flip) applied at different depths. We see that a phase-flip does not affect the output distribution if it is applied close to the end of the circuit. The reason is that we measure in the computational basis, which is insensitive to phase errors. Furthermore, the two-qubit $\CZ$ gates used in the circuit commute with $\Z$ errors.  

Figure.~\ref{fig:xcorr} shows the residual correlation for a single $\X$ error (bit-flip). Bit-flip errors do not have any effect after the cycle of Hadamards at the beginning of the circuit (see Sec.~\ref{sec:pt}), which rotate the initial state (in the computational basis) to the $x$ basis. Some bit-flip errors towards the end of the circuit also do not affect correlations because the corresponding $\X$ error can get acted upon by a Hadamard-like gate, such as $\Y^{1/2}$. This rotates the $\X$ error into the $z$ basis, in which the state is measured.

\section{Quantum Simulation Details}~\label{app:qsd} 
In this appendix we summarize the implementation, optimization and performance of our high-performance gate-level quantum simulator. Additional details are available in ~\cite{2016arXiv160107195S,2016arXiv160406460H}. This simulation was used for all the circuits with $6\times6$ and $7\times6$ qubits. Simulations of smaller circuits, including all the simulations with errors, were performed with a different simulator running in local workstations. 

In order to simulate  quantum circuits on a classical computer, we implement a distributed high-performance quantum simulator that can simulate general single-qubit gates and two-qubit controlled gates. We perform a number of single- and multi-node optimizations, including vectorization, multi-threading, cache blocking, as well as gate specialization to avoid communication. Using Edison, distributed Cray XC30 system at National Energy Research Scientific Computing Center (NERSC), we simulate random quantum circuits of up to 42 qubits, with an average time per gate of 1.72
 seconds. These are the largest quantum circuits simulated to-date for a computational task that approaches quantum supremacy.

\subsection{Background}
\label{subsec: background}

Given $n$ qubits, our simulator evolves a $2^n$ state vector, using single-qubit as well as two-qubit controlled gates. Let $U_{\rm sq}$ be a $2\times 2$ unitary matrix that represents a single-qubit gate operation:
\[ 
U_{\rm sq}=\left(
\begin{array}{cc}
u_{11} & u_{12} \\
u_{21} & u_{22} \\
\end{array} 
\right)\;.
\]
To perform gate $U_{\rm sq}$ on qubit $k$ of the $n$-qubit quantum register, we apply $U_{\rm sq}$ to the pairs of amplitudes whose indices differ in the $k$-th bits of their binary index:
\begin{equation}
	\begin{aligned}
		\alpha'_{*...*0_{k}*...*} = u_{11} \cdot \alpha_{*...*0_{k}*...*} + u_{12} \cdot \alpha_{*...*1_{k}*...*} \\
		\alpha'_{*...*1_{k}*...*} = u_{21} \cdot \alpha_{*...*0_{k}*...*} + u_{22} \cdot \alpha_{*...*1_{k}*...*}  
		\label{eq:sqgunitaryoperation}
	\end{aligned}	
\end{equation}

A generalized two-qubit controlled-$U$ gate, with  a control qubit $c$ and a target qubit $t$, works similarly to a single-qubit gate, except that only the pairs of amplitudes for which $c$ is set are affected, while all other amplitudes are left unmodified.

\subsection{Implementation and Optimization}
\label{sec:implandopt}

The implementation of single- and two-qubit controlled gates  follows directly  Eq.~\ref{eq:sqgunitaryoperation}. For example, to apply a single-qubit gate to qubit $k$, we iterate over consecutive groups of amplitudes of length $2^{k+1}$,  applying $U_{\rm sq}$ to every pair of amplitudes that are $2^k$ elements apart. To achieve high performance, we perform the following optimizations.

\textit{Vectorization:} Exploring data parallelism is fundamental to the high performance and energy efficiency of modern architectures. Modern Intel CPUs support data parallelism in the form of SIMD (Single Instruction Multiple Data) instructions, such as AVX2~\cite{AVX2}. These instructions perform four double-precision operations simultaneously on four elements of the input registers. Our implementation maps every two pairs of complex amplitudes into four-wide SIMD instructions; each pair, which operates on real and imaginary parts, uses half of the SIMD register.\footnote{Intel recently announced that the second generation $\text{Intel}^{\textregistered}$ Xeon $\text{Phi}^{\texttrademark}$ architecture will also support eight-wide AVX512. This will allow simultaneous operations on four pairs of amplitudes, and will enable additional performance benefits.}

\textit{Multithreading:} Modern multi- and many-core CPUs support execution of many concurrent hardware threads.   We parallelize single- and two-qubit controlled  gate operations on these threads using OpenMP 4.0~\cite{openmp13}. We adaptively exploit thread-level parallelism either across groups or within a single group. Namely, we first try to divide groups of amplitudes evenly among all threads. When there are not enough groups to use all available threads, we explore thread parallelism within a group.

\textit{Cache Blocking:} Single and controlled qubit operations perform a small amount of computation, and, as a result, their performance is limited by memory bandwidth. To increase arithmetic intensity of the quantum simulator, one can form larger gate matrices as a tensor product of several parallel gates. As a result,  subsets of amplitudes are reused over matrix columns, but at the expense of redundant computation, which grows exponentially with the number of combined gates.  Our approach identifies and operates on groups of consecutive gates which update a small portion of the state vector, common to all the gates, that also fits into Last Level Cache (LLC). LLC offers much higher bandwidth than main memory, which improves the performance of the simulator. LLC also has much smaller capacity, which limits this optimization only to the gates that operate on lower-order qubits~\cite{2016arXiv160107195S}.

\textit{Multi-node Implementation:} Single node quantum simulation is limited by the size of the physical memory of the compute node.\footnote{While is conceivable to hold the state on the secondary storage device, the latter is significantly slower than main memory, thus rendering most interesting quantum simulations unpractical.} To simulate larger numbers of qubits requires a distributed implementation. Our distributed simulation partitions a state vector of $2^n$ amplitudes ($2^{n+4}$ bytes) among $2^p$ nodes, such that each node stores a local state of $2^{n-p}$ amplitudes.   Given single- or controlled two-qubit  gate operations on the target qubit $k$, if $k < n-p$, the operation is fully contained within a node; otherwise it requires inter-node communication. Our communication scheme follows~\cite{DBLP:phd/de/Trieu2010}, where two nodes exchange half of their state vectors into each other's temporary storage, compute on exchanged halves, followed by another pair-wise exchange. In contrast to ~\cite{DBLP:phd/de/Trieu2010} which requires large temporary space to hold exchanged halves, our implementation requires very small temporary storage and is thus much more memory efficient.

\textit{Gate Specialization~\cite{2016arXiv160406460H,personalcomm}.} To further reduce the run-time of the simulator, we take advantage of the specialized structure of each gate matrix. For example, the entries of a Hadamard matrix are real, which reduces the extra overhead of complex arithmetic. This is particularly helpful when combined with cache blocking which makes the simulation more compute bound.  Recognizing diagonal gates, such as  $\T$ gates, allows one to avoid inter-node communication, while recognizing an entry equal to $1.0$ on the main diagonal of the diagonal gates (as in $\Z$ or $\T$ gates), reduces memory bandwidth requirements by $2\times$, and results in commensurate performance improvements.

\subsection{Performance}
\label{sec: performance}

We performed quantum simulations on Edison supercomputer~\cite{Edison}. Edison is a distributed Cray XC30 system at National Energy Research Scientific Computing Center (NERSC), ranks \# 39 in the latest TOP500 list, and consists of 5,576 compute nodes. Each node is a dual-socket Intel\textregistered Xeon E5 2695-V2 processor with 12 cores per socket, each running at 2.4GHz. Each core is a superscalar, out-of-order core that supports 2-way hyperthreading and offers AVX support. All 12 cores share a 30MB L3 last level cache and a memory controller connected to four DDR3-1600 DIMMs that together provide 64GB of memory per node (32GB per socket). The nodes are connected via Cray Aries with Dragonfly topology.  We use OpenMP~4.0~\cite{openmp13} to parallelize computation among threads. We also use  $\text{Intel}^{\textregistered}$ Compiler v15.0.1 and $\text{Intel}^{\textregistered}$  Cray MPI 7.3.1 library.

The time to simulate an \mbox{$n$-qubit} quantum circuit on $2^p$ nodes is proportionate to
\begin{equation*}
f \frac{G 2^{n-p}}{B_{\rm mem} } + (1-f) \left(\frac{G 2^{n-p}}{B_{\rm mem} } + \frac{G 2^{n-p}}{B_{\rm net}}  \right)\;.
\end{equation*}
Here, $G$ is the total number of gates,  $B_{\rm mem}$  is achievable  memory bandwidth,  $B_{\rm net}$ is achievable bidirectional network bandwidth, and $f$ is the fraction of  gates which do not require communication. The first term gives the time to simulate gates that do not require communication, while the second term gives the time to simulate gates that communicate. Thus we expect gate operations  which require  communication to be $1+B_{\rm mem}/B_{\rm net}$  slower than gates which communicate.
On Edison, the highest achievable memory bandwidth is $50$ GB/s per socket, while the highest achievable bidirectional network  bandwidth is $7$ GB/s per socket~\cite{Austin2013}. Thus the expected slowdown of gates that require communication, compared to gates that do not,  is $\sim 8{\times}$.

\begin{figure}
 	\centering
 	\includegraphics[width=\columnwidth]
 	{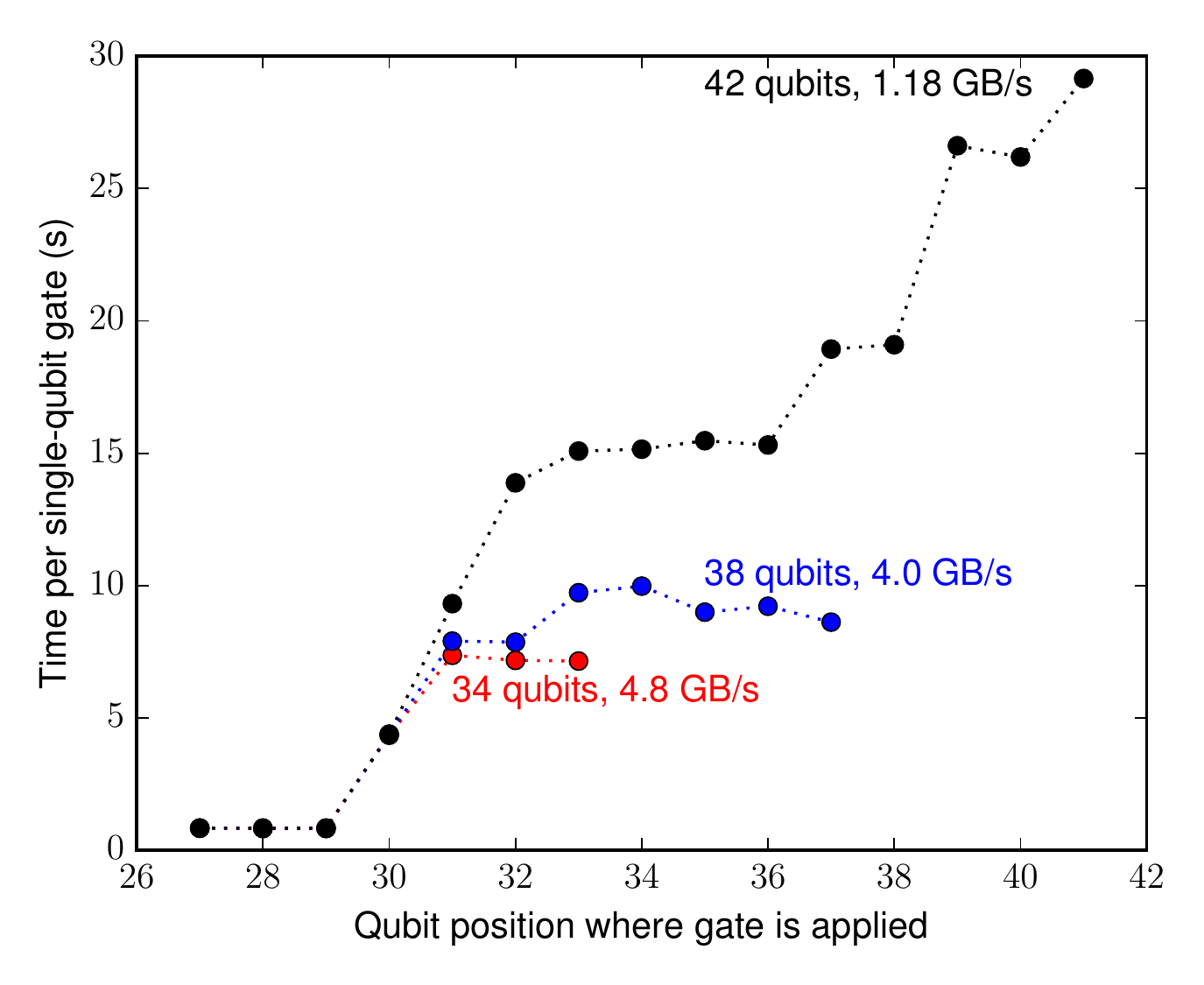}
 	\vspace{1mm}
 	\caption{Gate benchmarking results on multiple nodes (sockets) for the single-qubit Hadamard gate. The $x$-axis is the position of the qubit where the gate is applied. Operations on qubits in position 30 and above require network communication. The magnitude of the jump in the time per gate after position $30$ is commensurate with the ratio between network and memory bandwidth.  Numbers in the labels show achieved bandwidth for the higher ordered qubits.  }
 	\label{fig:weakscaling}
\end{figure}
 
Figure~\ref{fig:weakscaling} reports benchmarks of the performance of a single-qubit Hadamard gate on  16, 256, and 4,096 sockets, simulating 34, 38, and 42 qubits, respectively, while keeping the problem size per socket constant (i.e., $2^{30}$ double complex amplitudes, or $2^{34}$ bytes). Gates performed on qubits $0-30$ require no inter-socket communication and take ${\sim} 0.82$ seconds per gate. This corresponds to 42 GB/s memory bandwidth ($2 \text{ [accesses (read/write)]} \cdot 2^{34} \text{ [bytes]} \: / \: 0.82 \text{ [seconds]}$), or $84\%$ of highest achievable bandwidth. 

Gates applied to higher-order qubits, 30 and above, require communication, which increases the time per gate. For example, for a 36-qubit system simulated on 16 sockets, the time per gate increases to  7.6 seconds, which corresponds to 4.8 GB/s network bandwidth.  The $9{\times}$  increase compared to the no-communication case is consistent with our expectation, discussed earlier. As we increase the number of sockets, the time per gate further increases for higher order qubits. For example, for a 42-qubit system on 4,096 sockets, the time to apply a Hadamard gate to qubit 41 is 29 seconds -- a nearly three-fold increase compared to applying a Hadamard gate to qubit 31. This corresponds to 1.18 GB/s network bandwidth, which is almost a $6{\times}$ drop, compared to the best achievable bandwidth of 7 GB/s. This drop is consistent with the detailed bandwidth analysis of Aries interconnect in Ref.~\cite{Austin2013}. Intuitively, the drop is due to the fact that higher-order qubits result in a larger distance between communicating sockets, which, in turn, results in increased volume of communication over global links and thus strains the bi-section bandwidth of the system.

\begin{table*}[htbp]
	\centering
	\begin{tabular}{|r|c|c|c|c|c|}
		\hline
		\textbf{Optimization} &  \textbf{\% of } & \textbf{\# of} & \textbf{\# of } & \textbf{Avg. time} & \textbf{Time per} \\ 
		\textbf{Level} &  \textbf{comm} & \textbf{sockets} & \textbf{fused} & \textbf{per gate (sec)} & \textbf{Depth-25 (sec)} \\\hline

		\multicolumn{6}{|c|}{\textit{$5 \times 4$ circuit}: \textcolor{blue}{$20$ qubits, 10.3 gates per level, \textbf{17 MB of memory}}}  \\ \hline
		no spec		 	& 0.0\% 	& 1     	& n/a   	& 0.00022 	& 0.057	 \\
		spec  	 	& 0.0\% 	& 1     	& n/a   	& 0.00015 	& 0.039	 \\
		spec+cb  	& 0.0\% 	& 1     	& 0.00  	& 0.00015 	& \textbf{0.039} \\\hline
		\multicolumn{6}{|c|}{\textit{$6 \times 4$ circuit}: \textcolor{blue}{$24$ qubits, 12.5 gates per level, \textbf{268 MB of memory}}}  \\\hline
		no spec 	  	& 0.0\% 	& 1     	& n/a   	& 0.0111 	& 3.466  \\
		spec  		  	& 0.0\% 	& 1     	& n/a   	& 0.0088 	& 2.741	 \\
		spec+cb         & 0.0\% 	& 1     	& 7.01  	& 0.0041 	& \textbf{1.294} \\\hline
		\multicolumn{6}{|c|}{\textit{$6 \times 5$ circuit}: \textcolor{blue}{$30$ qubits, 16.2 gates per level, \textbf{17 GB of memory}}} \\\hline
		no spec 	 	& 0.0\% 	& 1     	& n/a   	& 0.721	  	& 292.2	 \\
		spec  		  	& 0.0\% 	& 1     	& n/a   	& 0.572	  	& 231.8  \\
		spec+cb      	& 0.0\% 	& 1     	& 5.64  	& 0.349	  	& \textbf{141.3} \\\hline
		\multicolumn{6}{|c|}{\textit{$6 \times 6$ circuit}: \textcolor{blue}{$36$ qubits, 19.5 gates per level, \textbf{1 TB of memory}}}  \\\hline
		no spec 	  	& 15.9\% 	& 32    	& n/a   	& 1.51  	& 735.1 \\
		spec  		 	& 6.2\% 	& 32    	& n/a   	& 1.08  	& 526.7 \\
		spec+cb   	& 6.2\% 	& 64    	& 5.40  	& 0.76  	& \textbf{369.0} \\\hline
		\multicolumn{6}{|c|}{\textit{$7 \times 6$ circuit}: \textcolor{blue}{$42$ qubits, 23.0 gates per level, \textbf{70 TB of memory}}}  \\\hline
		spec+cb    	& 11.2\% 	& 4,096  	& 5.54  	& 1.72  	& \textbf{989.0} \\ \hline
		
	\end{tabular}%
	\caption{Simulator performance comparison of five random circuits: \textit{$5 \times 4$},  \textit{$6 \times 4$}, \textit{$6 \times 5$}, \textit{$6 \times 6$}, and \textit{$7 \times 6$} . First column lists three levels of optimizations, for each circuit. Second column shows the fraction of gates which require communication ($1-f$). Third and fourth columns show the number of sockets used, and average number of fused gates to enable cache blocking ($cb$) optimization, respectively (see Sec.~\ref{sec:implandopt}). The last two columns show average time per gate and time per circuit with depth 25, respectively. }
	\label{table:configs}%
\end{table*}%

Table~\ref{table:configs} compares simulator performance characteristics of five random circuits with different lattice dimensions and number of qubits.  The table is broken into five sections, one for each circuit. For each circuit, we show the characteristics for three levels of optimization: without specialization, with specialization, and with both specialization and cache blocking ($cb$) enabled. Circuits with 20, 24 and 30 qubits are simulated on a single socket, while circuits with 36 and 42 qubits are simulated on 64 and 4,096 sockets, respectively.

Specializing the gates reduces run-time of a 20-qubit circuit by $1.46{\times}$, compared to $1.26{\times}$ run-time reduction for 24- and 30-qubit circuits, as shown in the first three sections of the table. As mentioned in Section~\ref{sec:implandopt},  specializing gates, such as $\T$ and $\CZ$, reduces  memory traffic by $2{\times}$. This reduces the simulation time of these gates on 24- and 30- qubit systems, whose state does not fit into Last Level Cache (LLC), making their performance  bounded by memory bandwidth. In addition to $2{\times}$ reduction in memory traffic, gate specialization also reduces compute requirements by as much as $4{\times}$: for example, without specialization, applying a $\T$ gate results in four complex multiply-adds per pair of state elements, while with specialization applying a $\T$ gate results in only one complex multiply-add. This reduces the simulation time of a 20-qubit system, whose 17 MB state fits into the 30 MB of the Last Level Cache (LLC), making its performance compute-bound. Thus gate specialization results in higher run-time reduction for a 20-qubit circuit than for 24- and 30-qubit circuits. Another consequence of the fact that the state of a 20-qubit circuit fits into LLC is that cache blocking optimization does not take effect. Furthermore, for 24- and 30-qubit circuits, cache blocking reduces the average time per gate by  $2.1{\times}$ and $1.6{\times}$, respectively. A 30-qubit circuit benefits less from cache blocking, compared to a 24-qubit circuit, because it has fewer gates that can be fused, as shown in the fifth column.

The last two sections of the table show performance statistics for a 36- and a 42-qubit circuits, which are simulated on 64 and 4,096 sockets, respectively. As Figure~\ref{fig:weakscaling} shows, for a 36-qubit simulation the time per gate varies  between 0.8 seconds (when there is no communication) and  8  seconds (when communication is required). Note that only $16\%$  of the gates require communication, as shown in the second column. As a result, we measure an average time of 1.5 seconds per gate, as shown in the fourth column of the table. Gate specialization more than halves the number of gates that require communication. This results in 1.08 seconds per gate: $1.4{\times}$ reduction of average time per gate, compared to no specialization. Combining cache blocking optimization with specialization  reduces the time per gate down to 0.76 seconds: an additional $1.4{\times}$ reduction compared to specialization only. As shown in the fourth column, for a 36-qubit circuit, we are able to fuse over five consecutive gates, on average. Overall, both gate specialization and cache blocking reduce the average time per gate as well as the total run-time of a circuit with  depth 25 (last column) by nearly $2{\times}$.

The last row shows the simulator performance on a 42-qubit random circuit when both gate specialization and cache blocking are used.  Compared to a 36-qubit random circuit, the number  of gates per level on a 42-qubit random circuit has increased by almost $20\%$. In addition, as the second column shows, the fraction of gates that requires communication has increased by almost $2{\times}$, while the time per gate has also increased, as shown in Figure~\ref{fig:weakscaling}.  As a result, the average time per gate on a 42-qubit simulation is 1.72 seconds; a $2.3{\times}$ increase compared to a 36-qubit simulation. Overall, it took 1,589 seconds  to simulate a 42-qubit circuit with the depth of 25: 989 seconds ($1.72 {\text{ seconds per gate}} \times 23.0 {\text{ gates per level}} \times 25 {\text{ levels}} $) to simulate all the gates, and 600 seconds to compute statistics, such as entropy, the cross entropy with the uniform distribution and probability moments.

An improved implementation of a quantum circuit simulator was recently reported in Ref.~\cite{haner2017petabyte} after this paper appeared in the arXiv. Ref.~\cite{haner2017petabyte} obtains an order of magnitude speedup against the benchmarking reported here for circuits with 42 qubits, and reports simulations of circuits with 45 qubits. Nevertheless, if as done in Figs.~\ref{fig:evd} and Fig.~\ref{fig:mvd}, we want to obtain statistics of the final state at each cycle of the quantum circuit for scientific purposes, the relative speedup will be substantially diminished. 

\section{Numerical estimation of the treewidth of the Ising model}\label{app:treewidth}
\begin{figure}
  \centering
  \includegraphics[width=\columnwidth]{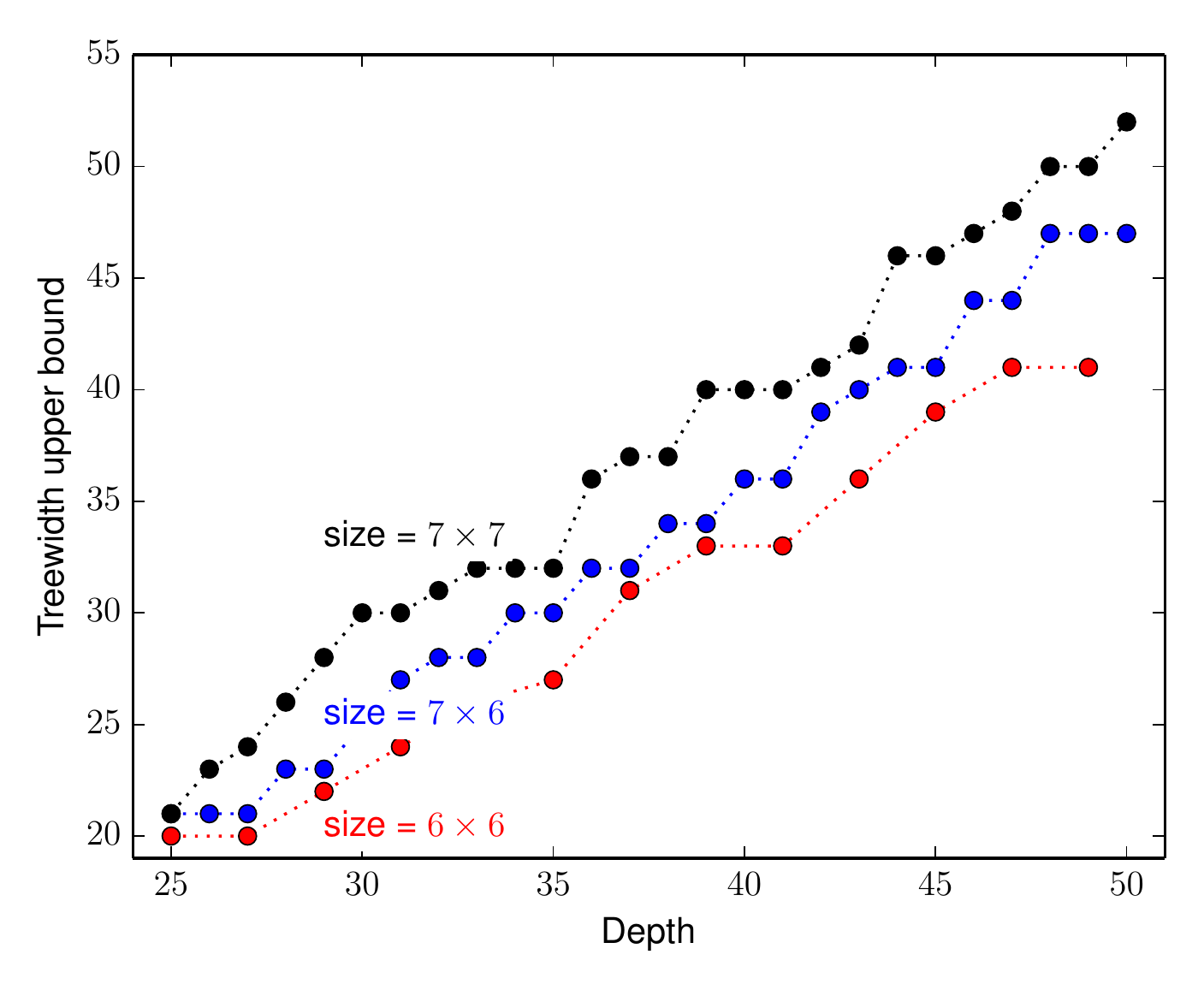}
  \caption{Numerical upper bound for the treewidth of the interaction graph of the Ising model corresponding to circuits with $6 \times 6$, $7 \times 6$, and $7 \times 7$ qubits as a function of the circuit depth (see Sec.~\ref{sec:pf}).}
  \label{fig:treewidth}
\end{figure}

For a circuit in a 2D lattice of qubits with two-qubit gates restricted to nearest neighbors, the treewidth of the corresponding Ising model (see Sec.~\ref{sec:ch}) is proportional to  $\min(d\sqrt n, n)$. Figure~\ref{fig:treewidth} shows numerical upper bounds for the treewidth as a function of depth for the circuits in Sec.~\ref{sec:pt}. The upper bounds were obtained by running the \emph{QuickBB} algorithm~\cite{gogate2004complete}. 

\section{Non-Clifford gates}\label{app:t_gates}
\begin{figure}
  \centering
  \includegraphics[width=\columnwidth]{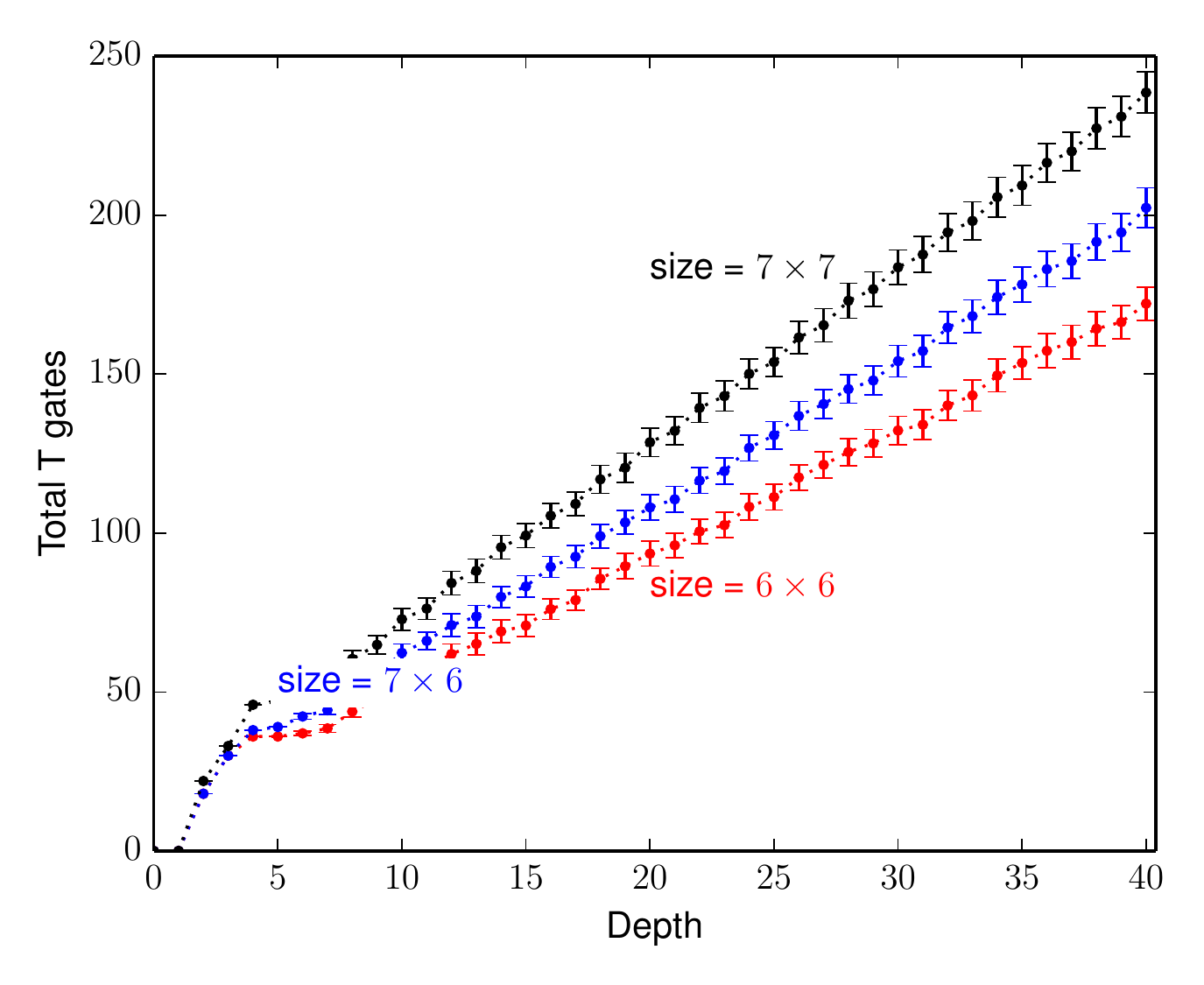}
  \caption{Number of non-Clifford $T$ gates as a function of depth for circuits with $6 \times 6$, $7 \times 6$, and $7 \times 7$ qubits. Error bars are the standard deviations among random circuit instances. }
  \label{fig:t_gates}
\end{figure}

Clifford circuits (circuits which only contain Clifford gates) can be simulated efficiently~\cite{gottesman1998heisenberg}. Furthermore, this method can be extended to simulate circuits which are dominated by Clifford gates~\cite{bravyi_improved_2016}. The only non-Clifford gate employed on the circuits we have used, as defined in Sec.~\ref{sec:pt}, is the $\T$ gate. Figure~\ref{fig:t_gates} plots the number of $T$ gates. On the one hand, the number of $T$ gates is likely too big for this simulation method to work for circuits with $7\times 7$ qubits and depth 40. This number can also be easily increased. On the other hand, the number of $T$ gates can be decreased at will, which will allow for the verification of circuits with even $7\times 7$ qubits, when a direct simulation is likely no longer possible. 

\section{Depth to reach Porter-Thomas for denser 2D circuits}\label{app:pt_nc}

\begin{figure}
  \centering\includegraphics[width=\columnwidth]{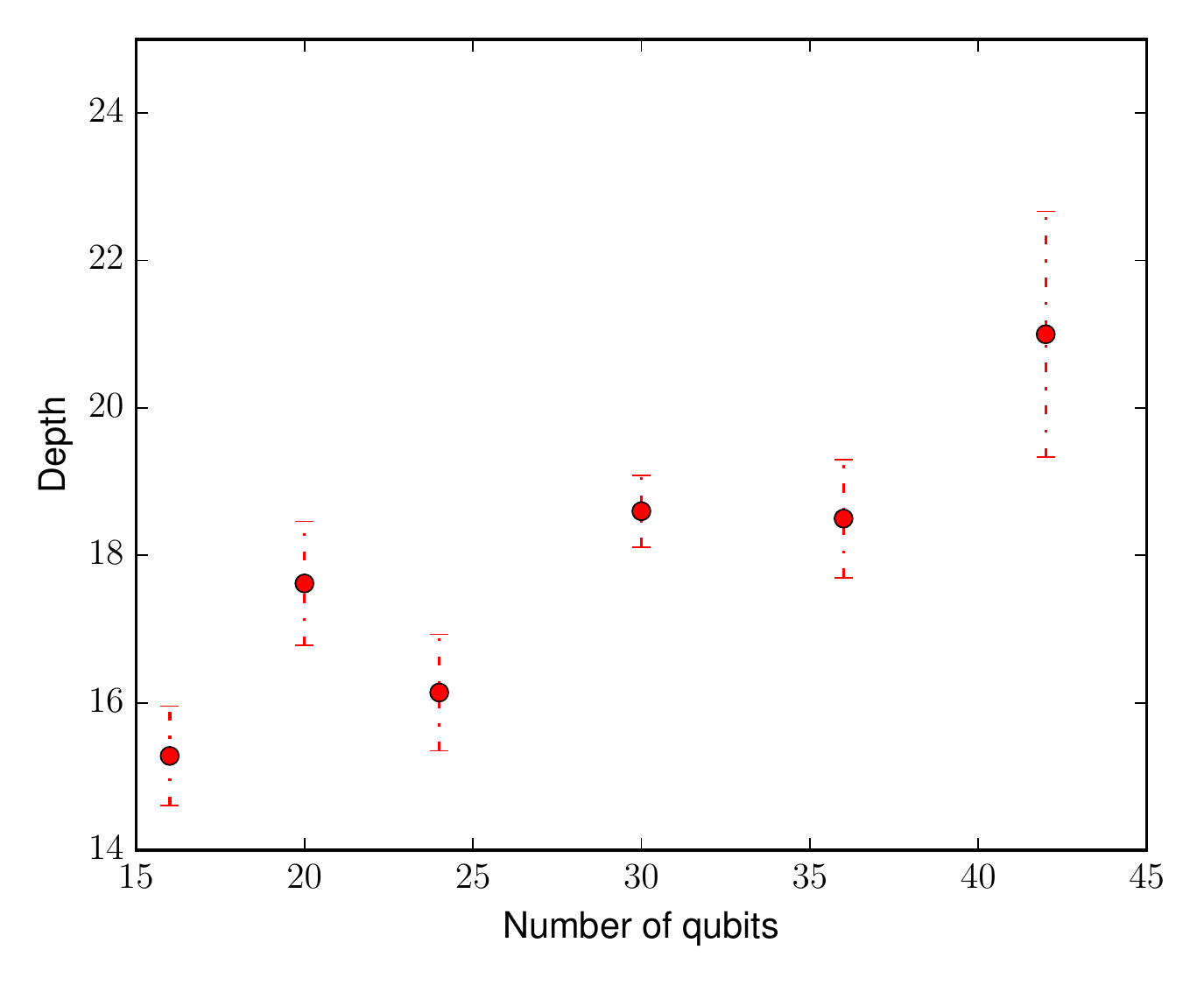}
  \caption{First cycle in a random circuit instance such that the entropy remains within $4$-sigma of the Porter-Thomas entropy during all the following cycles. Markers show the mean among instances and error bars correspond to the standard deviation among circuit instances.  Depth is measured in layers, and each layer is a cycle of random single-qubit followed by a cycle of $\CZ$ gates.}
  \label{fig:pt_depth_nc}
\end{figure}  

It is currently not possible to perform two CZ gates simultaneously in two neighboring superconducting qubits~\cite{barends_superconducting_2014,barends_digital_2015,kelly_state_2015,barends_digitized_2015}. This restriction was used for the circuits of the main text, see Fig.~\ref{fig:two_dimers}. In this appendix we report simulations of circuits in a 2D lattice where, ignoring this particular restriction, a two-qubit gate is applied to every qubit in each cycle of $\CZ$ gates.  In order to get a smoother scaling for circuits of different sizes, we use periodic boundary conditions for the layout of two-qubit gates. We find numerically a good convergence to the Porter-Thomas distribution for the following circuits.

We use the same single-qubit gates as in the main text, $\{\X^{1/2}, \Y^{1/2}, \T\}$. In addition we use two-qubit $\CZ$ gates. The circuits are:
\begin{enumerate}
\item Initialize in the state $\ket{0}^{\otimes n}$.
\item Apply a Hadamard gate to each qubit.
\item Apply a random circuit with a stack of depth $d$, where each layer has the following two clock cycles:
  \begin{enumerate}
  \item Apply a clock cycle of random single-qubit gates to all qubits.
    \item Apply a clock cycle of two-qubit $\CZ$ gates.
  \end{enumerate}
\end{enumerate}
We follow the same restrictions for the placement of single-qubit gates as in Sec.~\ref{sec:pt}. For the cycle of two-qubit gates, we follow a similar sequence to the layouts of Fig.~\ref{fig:two_dimers}, but now every qubit participates in 
exactly one $\CZ$ gate. In addition, as mentioned above, we use periodic boundary conditions.

Figure~\ref{fig:pt_depth_nc} shows the first layer of each random circuit instance for which the entropy remains within $4$-sigma of the Porter-Thomas entropy during all the following layers (similar to Fig.~\ref{fig:pt_depth}). Note that we now measure the depth in layers, and each layer consists of a cycle of single-qubit gates and a cycle of two-qubit gates. Physically, though, cycles of single-qubit gates are normally faster than cycles of two-qubit gates. 

\section{Outline of Stockmeyer Counting Theorem}\label{app:stockmeyer}
In this section we outline the main ideas behind the Stockmeyer Counting Theorem~\cite{stockmeyer1983complexity,trevisan_lecture_2004,goldreich_computational_2008}. As discussed in Sec.~\ref{sec:cca} an NP-oracle is a computational complexity theory construct that determines if a given equation
\begin{align}\label{eq:fzx}
  f(z) = x
\end{align}
has any solutions, see for example Eq.~\eqref{eq:13}. The function $f$ maps bit-strings to bit-strings and can be evaluated in polynomial time in the input size $n$. The  Stockmeyer Counting Theorem states that an NP-oracle also suffices to determine, with high probability, an approximation $\tilde{q}(x)$ to the number of solutions $q(x)$ of Eq.~\eqref{eq:fzx}
\begin{align}
  \label{eq:sm}
  |\tilde{q}(x) - q(x)| < q(x)/\poly(n) 
\end{align}
where $\poly(n)$ denotes any chosen polynomial in $n$. The main ingredient is the use of so-called hash functions, described below, to estimate if there are at least $2^k$ solutions. The result then follows by trying different values of $k \le n$. 

A hash function $h_{n,m}$ maps an $n$-bit-string to an $m$-bit-string with $n>m$. Let's consider the subset $T_h$ of bit-strings which are mapped to $0$ by $h$. Let $H_{n,m}$ be a sufficiently random family of hash functions (a pairwise independent family). Let $S $ be a subset of $n$-bit-strings of size $|S|$ sufficiently larger than $2^m$.  Because a random $h_{n,m} \in H_{n,m}$ selects a random $T_h$, the size of the subset $S \cap T_h$ is concentrated around its expectation value $|S|/2^m$~\cite{Impagliazzo:1989:PGO:73007.73009}.

Consider now the set $S \equiv \{z : f(z) = x\}$ of solutions $z$  to Eq.~\eqref{eq:fzx}. We can use a random family of hash functions to construct an algorithm that with finite probability of success, $3/4$ for example, can distinguish between $|S| > 2^{k}$ and $|S| \le 2^k$, where $k = m+5$. 
 This is done using a single NP-oracle call to check if there are a finite number of elements $S$ mapped to 0, 48 for example, by a random hash function from $H_{n,m}$. The probability of success can be amplified to $1-1/(4 \kappa)$ with $\kappa$ invocations of the NP-oracle.

\section{Multiplicative approximation to $|Z|^2$ from the Porter-Thomas distribution}\label{sec:maz}
We recall from the discussion in Sec.~\ref{sec:pf} that each output probability of a random quantum circuit $p_U(x)$ is proportional to the partition function of a complex Ising model. In this appendix we review why approximate sampling with constant variational distance from the output of random circuits implies a probabilistic multiplicative error approximation to such partition functions with an NP-oracle~\cite{bremner2015average,aaronson2011computational}. We follow the proof from Ref.~\cite{bremner2015average},  but use the Porter-Thomas distribution, instead of their anti-concentration bound. 

Let $q(x)$ denote the output probability of a classical sampling algorithm for a bit-string $x$ of our choice, and  $\tilde{q}(x)$ an approximation obtained using the  Stockmeyer Counting Theorem. From Eq.~\eqref{eq:sm} and the triangle inequality we obtain
\begin{multline}
  \label{eq:16}
  |\tilde{q}(x) - p(x)| \le \(1+1/\poly(n)\) |q(x) - p(x)| \\+ p(x)/\poly(n) \;.
\end{multline}

Let us suppose what we want to disprove: a classical sampling algorithm $A_\cl(U)$ with probabilities $q(x)$ and polynomial computational time in $n$ which achieves an $\epsilon$ approximation in the variational distance to the output of any given quantum random circuit
\begin{align}
  \label{eq:cl1}
  \sum_x |q(x) - p_U(x)| < \epsilon\;.
\end{align}
We will show that then $p_U(x)$ can be approximated using Stockmeyer Counting Theorem, which is conjectured to be impossible.

From Markov's inequality we have, for any $0<\delta<1$,
\begin{align}
  \label{eq:14}
  \Pr_x\(|q(x) - p_U(x)| \ge \frac \epsilon {2^n \delta} \) \le \delta
\end{align}
where $x$ is picked uniformly at random. Setting $\delta = 4 \epsilon$ we obtain
\begin{align}
  \label{eq:17}
   \Pr_x\(|q(x) - p_U(x)| \le \frac 1 {2^{n+2} } \) \ge 1 - 4 \epsilon\;.
\end{align}
Therefore, with probability $1-4\epsilon$, we have
\begin{align}
  \label{eq:18}
  |\tilde{q}(x) - p_U(x)| \le { 1+1/\poly(n) \over 2^{n+2}}+ p_U(x)/\poly(n)\;.
\end{align}

Set, for example, $\epsilon = (8 e)^{-1} \approx 0.046$. If, as found numerically in Sec.~\ref{sec:pt}, we assume that the output of $U$ has Porter-Thomas distribution, then
\begin{align}
  \label{eq:19}
  \Pr\( p_U(x) > 2^{-n}\) = 1/e\;.
\end{align}
Eqs.~\eqref{eq:18} and \eqref{eq:19} imply that $\tilde{q}(x)$ approximates $p_U(x)$ up to a multiplicative error $1/4+o(1)$ with probability at least $1/e-4\epsilon =  (2e)^{-1}$. A similar bound can be found using only the second moment of the Porter-Thomas distribution~\cite{bremner2015average}.

From  Sec.~\ref{sec:pf} we have that $p_U(x) = \lambda |Z|^2$ where $\lambda$ is a positive known constant and $Z$ is the partition function of a complex Ising model. Therefore, if $\tilde{q}(x)$ approximates $p_U(x)$ up to a multiplicative error $1/4+o(1)$, then $\tilde{q}(x)/\lambda$ approximates $|Z|^2$ up to the same multiplicative error.

\section{Bayesian estimation of output probabilities}\label{app:feynman}

In this appendix we study a polynomial classical algorithm for approximately sampling the output distribution of a circuit $U$. The sampling follows from on an approximation to the output probability $p_U(x)$ of a bit-string $x$. As has been discussed in Sec.~\ref{sec:pf}, the amplitudes of the output state  of a random quantum circuit $U$ can be written in the form of a Feynman path integral where each path is encoded in the assignment of the vector $s$ of Ising spins, and the phase associated with the path is given by the energy of an Ising model $H_x(s)$.  The approximation algorithm considered here is a Bayesian estimation of the output probability of a given bit-string after randomly sampling a large number Feynman paths. 

The output amplitudes of a random circuit are proportional to the partition function of a random  Ising model $H_x(s)$ at complex temperature,
\begin{align}
  \Psi = \braket x {\psi_d} =\frac{1 }{\sqrt{L}}\sum_{k=0}^{K-1} M_k e^{i \frac {2\pi} K  k}
\end{align}
where the $k$'s are different energies of the Ising model (mod $K$), $L = 2^G$, $M_k \sim 2^{G}$, and $G$ is the number of two-sparse gates. The prefactor is $1/\sqrt L$ given the explicit choice of two-sparse gates, see Eq.~\eqref{eq:path1}.

We can always attempt to approximate the amplitude $\Psi$ for circuits of any size by sampling a large number $Q$ of spins configurations $s$ in the partition function. We start by counting the number of configurations $Q_k$ for each phase $k \in [0 \isep K-1]$ using the Ising model $H_x(s)$. We can assume that $1 \ll Q_k \ll L$. For example, the number of spins configurations is $L \sim 2^{250}$ for circuits with $7 \times 6$ qubits and depth 25. We will use the prior distribution from Porter-Thomas to derive the posterior distribution $\Pr(\Psi|\{Q_k\})$. We will see that the result is equivalent to a circuit fidelity $\sim Q^2/(NL)$. For instance, even if we sample $Q=10^{18}$ spin configurations this will give a fidelity of approximately $\sim 10^{-52}$ for a circuit with $7 \times 6$ qubits and depth 25.

Define the probabilities of the different paths as $p_k = M_k/L$. The prior probability of an amplitude from the Porter-Thomas distribution is
\begin{align}
  \Pr(\Psi) &\propto \exp(- N \Psi \Psi^*) \\
&= \exp\left( - N L \sum _{k_1=0}^{K-1} \sum _{k_2=0}^{K-1} p_{k_1}p_{k_2}\cos \frac{2 \pi  }{K} (k_1-k_2)\right)\;.\nonumber
\end{align}
We want to write the probabilities $p_k$ in a basis $v^\alpha$ that diagonalizes the kernel $\cos \frac{2 \pi  }{K} (k_1-k_2)$,
\begin{align}
\sum _{j=0}^{K-1} \cos \left(\frac{2 \pi  }{K} (m-j)\right) v^\alpha_j =  \lambda _{\alpha } v^\alpha_m
\end{align}
for $\alpha\in[0 \isep  K-1]$. The components $v^\alpha_j$ of the eigenvectors of the kernel are
\begin{align}
  v^0_j & = \frac 1 {\sqrt{K}}  \\ 
   v^\alpha_j &= \sqrt{\frac{2}{K}} \cos \left(\frac{2 \pi  }{K}\alpha  j \right), \quad \alpha \in [1 \isep K/2-1] \\
  v^{K/2}_j & = \frac {(-1)^j} {\sqrt{K}} 
\end{align}
and
\begin{align}
   v^\alpha_j &= \sqrt{\frac{2}{K}} \sin \left(\frac{2 \pi  }{K}(K-\alpha ) j \right), \; \alpha \in [K /2+1 \isep K-1] \nonumber\;.
\end{align}
The eigenvalues are
\begin{align}
  \lambda _{\alpha }=\frac{K}{2}  \left(\delta _{\alpha ,1}+\delta _{\alpha ,K-1}\right)\;.
\end{align}

Let $c'_j$ be the components of the vector of probabilities $p_k$ in the basis $v^\alpha$. We renormalize them to $c_j = c'_j$ for $j \notin \{1,K-1\}$ and $c_{\{1,K-1\}} = \sqrt{NLK/2} \,c'_{\{1,K-1\}}$ to write
\begin{align}
  \Pr(\Psi) \propto \exp(-(c_1^2 + c_{K-1}^2))\label{eq:cpt}
\end{align}
and
\begin{multline}
  p_j=\frac{2 }{K}\sqrt{\frac{1}{L N}} \left(c_1 \cos \left(\frac{2 \pi }{K} j\right)+ c_{K-1} \sin \left(\frac{2 \pi }{K} j\right)\right)\\+ \frac{1}{K} + \sum _{\alpha =2}^{K-2} c_{\alpha } v^\alpha_j\;.\label{eq:pjv}
\end{multline}
We define
\begin{align}
  \rho _k\equiv \sum _{\alpha =2}^{K-2} c_{\alpha } v^\alpha_k\;.\label{eq:f10}
\end{align}
With this definition, the numbers $\rho_k$ obey the following constraints
\begin{multline}
  0 = \sum _{k=0}^{K-1} \rho _k \cos \left(\frac{2 \pi }{K} k\right)=\sum _{k=0}^{K-1} \rho _k \sin \left(\frac{2 \pi }{K} k\right)\\=\sum _{k=0}^{K-1} \rho _k\;,\label{eq:f11}
\end{multline}
which will be used later to simplify the posterior probability.

The posterior probability for $\Psi$ is
\begin{align}
  \Pr(\Psi|&\{Q_k\}) \propto Q! \prod _{k=0}^{K-1} \frac{p_k^{Q_k}}{Q_k!} \Pr(\Psi) \\
 & \propto \exp \left(\sum _{k=0}^{K-1}   Q_k\log p_k\right)\exp \left(-\left(c_1^2+c_{K-1}^2\right)\right) \;.\nonumber
\end{align}
The log posterior  for $c_1, c_{K-1}$ is
\begin{multline}
  \log  \Pr\left(c_1,c_{K-1},\{\rho_j\} |\{Q_k\}\right) \\ \propto \sum _{j=0}^{K-1} Q_j \log \left( p_j\right)-\left(c_1^2+c_{K-1}^2\right)\;.\label{eq:lp1v}
\end{multline}

We are interested in the posterior probability $p = |\Psi|^2$. Note that $N p = c_1^2 + c_{K-1}^2$, as seen in the Porter-Thomas form of Eq.~\eqref{eq:cpt}. Therefore
\begin{multline}
  \Pr(p,\{\rho_j\}|\{Q\} )=\int _{-\infty }^{\infty }\int _{-\infty }^{\infty }  \Pr\left(c_1,c_{K-1},\rho |Q\right) \\ \delta\left(\frac{c_1^2+c_{K-1}^2}{N}-p\right) dc_1 dc_2 \;.
\end{multline}
After a change of variables $c_1=r \cos \phi$, $c_{K-1}=r \sin \phi$ we obtain
\begin{multline}
  \Pr(p,\{\rho_j\}|\{Q_j\})= \\ \frac{ N}{2  }\int _0^{2 \pi }  \Pr\left( \sqrt{N p}  \cos\phi , \sqrt{N p} \sin \phi ,\rho |\{Q\}\right)d\phi\;.\label{eq:f15}
\end{multline}

Using Eq.~\eqref{eq:pjv} in Eq.~\eqref{eq:lp1v} we write the Taylor series for the log posterior for $c_1, c_{K-1}$ as
\begin{multline}
  \log  \Pr \left(  \sqrt{N p} \cos \phi, \sqrt{N p} \sin  \phi,\{\rho_j\} |\{Q_k\}\right)\propto \\  \sum _{q=1}^{\infty } \frac{(-1)^{q+1}}{q L^{q/2}}  p^{q/2} \sum _{j=0}^{K-1} Q_j \left(\frac{2 \cos \left(\frac{2 \pi }{K} j-\phi \right)}{K \rho _j+1}\right)^q\\+
\sum _{j=0}^{K-1} Q_j \log \left(\rho _j+\frac{1}{K}\right)
-N p \;.\label{eq:lpts}
\end{multline}
We keep only the first term in $Q/L$ (using $1 \ll Q \ll L$), which is 
\begin{multline}
  \log  \Pr \left(  \sqrt{N p} \cos \phi, \sqrt{N p} \sin  \phi,\{\rho_j\} |\{Q_k\}\right)\propto \\   \sqrt{\frac p L}\sum _{j=0}^{K-1} Q_j \frac{2 \cos \left(\frac{2 \pi }{K} j-\phi \right)}{K \rho _j+1}\\+
\sum _{j=0}^{K-1} Q_j \log \left(\rho _j+\frac{1}{K}\right)
-N p \;.
\end{multline}
Exponentiating we get the posterior distribution
\begin{multline}
  \Pr \left(  \sqrt{N p} \cos \phi, \sqrt{N p} \sin  \phi,\{\rho_j\} |\{Q_k\}\right)\propto \\ 
  e^{-N p} \exp\left(\sum _{j=0}^{K-1} Q_j \log \left(\rho _j+\frac{1}{K}\right)\right) \\
  \Bigg( 1 + \sqrt{\frac p L}\sum _{j=0}^{K-1} Q_j \frac{2 \cos \left(\frac{2 \pi }{K} j-\phi \right)}{K \rho _j+1} \\ + \frac{2 p}{L} \left(\sum _{j=0}^{K-1}Q_j \frac{ \cos \left(\frac{2 \pi }{K} j-\phi \right)}{K \rho _j+1}\right)^2\Bigg)\;.\label{eq:f18}
\end{multline}
Note that we keep the second term when exponentiating, which is order $Q^2/L$, but we drop the second term in Eq.~\eqref{eq:lpts}, which is of order $Q/L^{3/2}$. 

We can carry out a further simplification by noticing that  $\rho_j$, which is defined in Eq.~\eqref{eq:f10} from the vector of probabilities $p_j$, obeys $\rho_j \ll Q$. Therefore, from the form of Eq.~\eqref{eq:f18}, we see that $\Pr(p |\{Q_k\}) \simeq \Pr(p,\{\bar \rho_j\} |\{Q_k\})$, where $\bar \rho_j$ is the expectation value of $\rho_j$ consistent with $\{Q_k\}$. This value can be obtained maximizing the posterior Eq.~\eqref{eq:lp1v} subject to the constraints given in Eq.~\eqref{eq:f11}.

We now insert Eq.~\eqref{eq:f18} into Eq.~\eqref{eq:f15} and carry out the integration to obtain
\begin{multline}
  \Pr(p|\{Q_k\}) = C e^{-Np} \exp \( \sum _{j=0}^{K-1} Q_j \log \left(\bar{\rho }_j+\frac{1}{K}\right)\) \\  
\Bigg( 1 + \frac p L \sum_{j_1,j_2=0}^{K-1} Q_{j_1}Q_{j_2} \frac{\cos \left(\frac{2 \pi  (j_1-j_2)}{K}\right)}{\left(K \bar \rho _{j_1}+1\right) \left(K \bar \rho _{j_2}+1\right)} \Bigg)\;.\label{eq:pbayes}
\end{multline}

Equation~\eqref{eq:pbayes} is the posterior probability $\Pr(p|\{Q_k\})$ for an approximation of the output probability $p=p_U(x)$ of a bit-string $x$ after sampling a large number $Q$ of spin configurations or Feynman paths in the expression for $\Psi$. We see that the probability $p$ enters explicitly in the last term, which is of the order $Q^2/L$. Next, we interpret this equation more formally. 

We argued in the text that the output state $\rho_\cK$ of an implementation with fidelity $\alpha$ of a quantum circuit $U$ can be modeled with Eq.~\eqref{eq:ra}
\begin{align}
  \rho_\cK = \alpha \ket{\psi_d}\bra{\psi_d} + (1-\alpha) \frac \openone N\;\label{eq:ra2}.
\end{align}
Then, the probability $p_U(x)$ for bit-strings $x$ sampled from an implementation with fidelity $\alpha$ has a distribution
\begin{align}
  \Pr_\alpha(p_U(x)) = N^2 e^{-Np}\(\alpha p + \frac {1-\alpha} N\)\;,\label{eq:f21}
\end{align}
see also Eq.~\eqref{eq:pal}. We can compare the posterior distribution, given by Eq.~\eqref{eq:pbayes}, with Eq.~\eqref{eq:f21} to obtain an equivalent ``fidelity'' $\alpha$ for the Bayesian classical approximate sampling algorithm, $\Pr(p|\{Q_k\})=\Pr_\alpha(p_U(x))$. 

First we obtain an expression for the normalization constant $C$ from the $p$-independent equation
\begin{align}
  C\, \exp \( \sum _{j=0}^{K-1} Q_j \log \left(\bar{\rho }_j+\frac{1}{K}\right)\)=(1-\alpha ) N\;.
\end{align}
The equation linear in $p$ gives
\begin{multline}
  N^2 \alpha = C \, \exp \( \sum _{j=0}^{K-1} Q_j \log \left(\bar{\rho }_j+\frac{1}{K}\right)\) \\
  \frac 1 L \sum_{j_1,j_2=0}^{K-1} Q_{j_1}Q_{j_2} \frac{\cos \left(\frac{2 \pi  (j_1-j_2)}{K}\right)}{\left(K \bar \rho _{j_1}+1\right) \left(K \bar \rho _{j_2}+1\right)}\;.
\end{multline}
Solving for $\alpha$ we obtain
\begin{align}
  \alpha = \frac 1 {NL} \sum_{j_1,j_2=0}^{K-1} Q_{j_1}Q_{j_2} \frac{\cos \left(\frac{2 \pi  (j_1-j_2)}{K}\right)}{\left(K \bar \rho _{j_1}+1\right) \left(K \bar \rho _{j_2}+1\right)}\;.
\end{align}
This is the final result, which shows that the equivalent circuit fidelity of the approximate sampling algorithm is $\alpha \sim Q^2/NL$, as promised.

\bibliographystyle{apsrev4-1} 
\bibliography{random_circuits}

\end{document}